\documentclass[a4paper, 11pt]{article}
\usepackage[pdftex, a4paper]{geometry}
\usepackage{graphicx}
\usepackage{subcaption}

\usepackage{amsmath}
\usepackage{amsfonts}
\usepackage{dsfont}
\usepackage{mathrsfs}
\usepackage{amssymb}
\usepackage{bbm}
\usepackage{epsfig}
\usepackage[english]{babel}
\usepackage[all]{xy}
\usepackage{color}

\newcommand*\mcapinn[2]{\vcenter{\hbox{$\mathsurround=0pt
  \ifx\displaystyle#1\textstyle\else#1\fi\bigcap$}}}

\newcommand*\mcup{\mathbin{\mathpalette\mcupinn\relax}}
\newcommand*\mcupinn[2]{\vcenter{\hbox{$\mathsurround=0pt
  \ifx\displaystyle#1\textstyle\else#1\fi\bigcup$}}}

\DeclareFontFamily{OT1}{pzc}{}
\DeclareFontShape{OT1}{pzc}{m}{it}{<-> s * [1.200] pzcmi7t}{}
\DeclareMathAlphabet{\mathpzc}{OT1}{pzc}{m}{it}

\newtheorem{theorem}{Theorem}
\newtheorem{definition}{Definition}

\newtheorem{lemma}{Lemma}

\newtheorem{proposition}{Proposition}

\title{\LARGE \bf Dynamics over Signed Networks\footnote{C.A. work was supported by the  Swedish Research Council (grant no. 2015-04390). A preliminary version of this paper was presented at the  56th IEEE Conference on Decision and Control, Melbourne,  Dec. 2017 \cite{cdc}.}}

\author{Guodong Shi\thanks{G. Shi is with the Australian Center for Field Robotics, 
School of Aerospace, Mechanical and Mechatronic Engineering,
The University of Sydney,  NSW 2006, Sydney, and the Research School of Engineering, The Australian National University, ACT 0200,
Canberra, Australia. Email: guodong.shi@sydney.edu.au}
, Claudio Altafini\thanks{C. Altafini is  with the Division of Automatic Control, Department of Electrical Engineering, Link\"{o}ping University, SE-58183 Link\"{o}ping, Sweden.
		E-mail:   claudio.altafini@liu.se}, and John S. Baras\thanks{J. S. Baras is  with the Institute for Systems Research and the Department of Electrical and Computer Engineering, University of Maryland, College Park, MD 20742, USA. E-mail: baras@isr.umd.edu}
}
\date{}

\begin{document}

\maketitle

\begin{abstract}
 A signed network is a network with each link  associated with a  positive or negative sign. Models for  nodes interacting over  such signed networks, where two different types of interactions take place along the positive and negative links, respectively, arise from various biological, social, political, and economic systems. As modifications to the conventional  DeGroot dynamics for positive links,  two basic types of    negative interactions along negative links, namely the opposing rule and the    repelling   rule, have been proposed and   studied in the literature. This paper reviews a few fundamental convergence results       for such dynamics over deterministic or random   signed networks under a unified   algebraic-graphical method. We show that    a systematic tool of studying node state evolution  over signed networks can be obtained utilizing generalized Perron-Frobenius theory, graph theory, and elementary algebraic recursions.
\end{abstract}

\section{Introduction}

In the past decades, the study of  network dynamics  has attracted tremendous  research  attentions from a variety of scientific disciplines  \cite{David-2010}. Particularly, with roots traceable back to topics such as 1960s products of stochastic matrices \cite{Wolfowitz-1963}, 1970s DeGroot social interactions models \cite{Degroot_1974}, and 1980s distributed optimization \cite{Tsitsiklis-Thesis}, consensus algorithms serve as a primary model for social network dynamics as well as being a foundation for some prominent engineering applications of large-scale complex networks \cite{Jadbabaie2003,Ren2005,Olfati-Saber-2007,Kar-TSP-2010,Golub-naivelearning-2010}. 

It has become a common understanding that cooperative node dynamics lead to  the emergence of  collective network behaviors.
On the other hand,  in various biological, social, political, and economical systems, there are often two different types of node interactions: activatory or  inhibitory,  trustful or mistrustful, cooperative or antagonistic \cite{Claudio-PNAS-2011,Kleinberg-PNAS-2011,Altafini_PONE12}. Using a positive or negative sign to represent  the type of a link, the structure of  these systems can be modeled as signed graphs.  After specifying the node dynamical relations along the positive or negative links, the evolution of node states defines the signed network dynamics. Consensus algorithms with positive and negative links have been recently proposed and  investigated \cite{Altafini_TAC13,Shi-JSAC13-2,Shi-TCNS15,Shi-TCNS17,Liu-Basar-15,Meng-SICON-2015,Meng-arXiv-2014,Shi-OR,Xia-Cao-TCNS-15,He14}. There exist   two basic types of    interactions along the negative links: the opposing  negative dynamics  \cite{Altafini_TAC13} where nodes are attracted by the opposite values of the neighbors,  and the repelling  negative dynamics   \cite{Shi-JSAC13-2} where nodes tend to be repulsive of the relative position of the states with respect to the neighbors.

\subsection{Signed Graphs}
Consider a network with $n$ nodes indexed in the set $\mathrm{V}=\{1,\dots,n\}$. The structure of the network is represented as an undirected graph $\mathrm{G}=(\mathrm{V},\mathrm{E})$, where an edge (link) $\{i,j\}\in\mathrm{E}$ is an unordered pair of two distinct nodes in the set $\mathrm{V}$. Each edge in $\mathrm{E}$ is associated with a sign, positive or negative, defining $\mathrm{G}$ as a signed graph. The positive and negative edges are collected in the sets $\mathrm{E}^+$ and $\mathrm{E}^-$, respectively. Then $\mathrm{G}^+=(\mathrm{V},\mathrm{E}^+)$ and $\mathrm{G}^-=(\mathrm{V},\mathrm{E}^-)$ are respectively termed positive and negative subgraphs. Throughout the paper and without further specific mention  we assume that $\mathrm{G}$ is connected and $\mathrm{G}^-$ contains at least one edge.

 For a node $i\in\mathrm{V}$, its positive neighbors are the nodes that share a positive link with $i$, forming  the set  $\mathrm{N}_i^+:=\big\{j:\{i,j\}\in\mathrm{E}^+\big\}$. Similarly the negative neighbor set of node $i$ is denoted as $\mathrm{N}_i^-:=\big\{j:\{i,j\}\in\mathrm{E}^-\big\}$. The set $\mathrm{N}_i=\mathrm{N}_i^+\mcup \mathrm{N}_i^-$ then contains all nodes that interact with node $i$ over the graph $\mathrm{G}$. We use ${\rm deg}_i=\big| \mathrm{N}_i\big|$ to denote the degree of node $i$, i.e., the number of neighbors of node $i$. Similarly, ${\rm deg}_i^+=\big| \mathrm{N}_i^+\big|$ and ${\rm deg}_i^-=\big| \mathrm{N}_i^-\big|$ represent the positive and negative degree of node $i$, respectively.

\subsection{Signed Laplacian}
   Let ${D}_{_{\mathrm{G}^+}}={\rm diag}({\rm deg}^+_1,\dots,{\rm deg}^+_n)$ and ${D}_{_{\mathrm{G}^-}}={\rm diag}({\rm deg}^-_1,\dots,{\rm deg}^-_n)$ be the  degree matrix of the positive subgraph and negative subgraph, respectively. Let ${A}_{_{\mathrm{G}^+}}$ be the adjacency matrix of the graph $\mathrm{G}^+$ with $[{A}_{_{\mathrm{G}^+}}]_{ij}=1$ if $\{i,j\}\in\mathrm{E}^+$ and $[{A}_{_{\mathrm{G}^+}}]_{ij}=0$ otherwise. The adjacency matrix ${A}_{_{\mathrm{G}^-}}$ of the negative subgraph $\mathrm{G}^-$ is defined by $[{A}_{_{\mathrm{G}^-}}]_{ij}=-1$ for $\{i,j\}\in\mathrm{E}^-$ and $[{A}_{_{\mathrm{G}^-}}]_{ij}=0$ for $\{i,j\}\notin\mathrm{E}^-$.

The Laplacian plays a central role in the algebraic representation of structural 
properties for graphs    \cite{Magnus-book}.  In presence of negative edges, more than one definition of Laplacian is possible,  e.g., \cite{Altafini_TAC13,Altafini_TAC15,siam-2014-spectrum}.
The  Laplacian of the positive  subgraph $\mathrm{G}^+$ is 
$
{L}_{_{\mathrm{G}^+}}:={D}_{_{\mathrm{G}^+}}-{A}_{_{\mathrm{G}^+}} 
$,
while for the negative subgraph $\mathrm{G}^-$ the following two variants can be used:
$
{L}_{_{\mathrm{G}^-}}^{^{\rm o}}:={D}_{_{\mathrm{G}^-}}-{A}_{_{\mathrm{G}^-}}
$
and
$
{L}_{_{\mathrm{G}^-}}^{^{\rm r}}:=-{D}_{_{\mathrm{G}^-}}-{A}_{_{\mathrm{G}^-}}.
$
Consequently, we have the following  definitions.
\begin{definition}
Given the signed graph $\mathrm{G}$, its opposing Laplacian is defined as 
\begin{equation}
{L}_{_{\mathrm{G}}}^{^{\rm o}}:={L}_{_{\mathrm{G}^+}}+{L}_{_{\mathrm{G}^-}}^{^{\rm o}} 
={D}_{_{\mathrm{G}^+}} + {D}_{_{\mathrm{G}^-}}-{A}_{_{\mathrm{G}^+}}  -{A}_{_{\mathrm{G}^-}},
\label{eq:laplacian-o}
\end{equation}
and its repelling Laplacian is defined as
\begin{equation}
{L}_{_{\mathrm{G}}}^{^{\rm r}}={L}_{_{\mathrm{G}^+}}+{L}_{_{\mathrm{G}^-}}^{^{\rm r}}
:={D}_{_{\mathrm{G}^+}} - {D}_{_{\mathrm{G}^-}}-{A}_{_{\mathrm{G}^+}}  -{A}_{_{\mathrm{G}^-}}.
\label{eq:laplacian-r}
\end{equation}
\end{definition}
The two superindexes ``o'' and ``r'' stand for ``opposing'' and ``repelling'' rules, terminology which will be introduced in Section~\ref{sec:pos-neg-int} and used throughout the paper\footnote{We prefer to avoid ambiguous terms like ``signed Laplacian'', which have been used in the literature to indicate both $ {L}_{_{\mathrm{G}}}^{^{\rm o}} $ and $ {L}_{_{\mathrm{G}}}^{^{\rm r}}$.}.
The two Laplacians $ {L}_{_{\mathrm{G}}}^{^{\rm o}} $ and $ {L}_{_{\mathrm{G}}}^{^{\rm r}}$ have different properties. For instance, $ {L}_{_{\mathrm{G}}}^{^{\rm o}} $ is always diagonally dominant, while $ {L}_{_{\mathrm{G}}}^{^{\rm r}}$ may or may not be; $ {L}_{_{\mathrm{G}}}^{^{\rm r}}$ has always zero as an eigenvalue, while $ {L}_{_{\mathrm{G}}}^{^{\rm o}} $ may or may not have it. 
Denote $\mathbf{x}=(x_1 \ \dots \ x_n)^\top$. Then we have the two induced  quadratic forms:
\begin{align}
\mathbf{x}^\top{L}_{_{\mathrm{G}}}^{^{\rm o}} \mathbf{x}&= \sum_{\{i,j\}\in \mathrm{E}^+} (x_i-x_j)^2+ \sum_{\{i,j\}\in \mathrm{E}^-} (x_i+x_j)^2, \label{lq1} \\
\mathbf{x}^\top{L}_{_{\mathrm{G}}}^{^{\rm r}} \mathbf{x}&= \sum_{\{i,j\}\in \mathrm{E}^+} (x_i-x_j)^2- \sum_{\{i,j\}\in \mathrm{E}^-} (x_i-x_j)^2. \label{lq2}
\end{align}
The two definitions \eqref{eq:laplacian-o} and \eqref{eq:laplacian-r} can be straightforwardly generalized the the weighted sign graph case where each link is associated with a positive or negative real number as its weight.

\subsection{Structural Balance Theory}
Introduced in the 1940s \cite{Heider-1946} and primarily motivated by social-interpersonal and economic networks, a fundamental notion in the study of signed graphs is the so-called {structural balance}. We recall the following definition (see  \cite{David-2010} for a detailed introduction).

\begin{definition}\label{definition-structural-balance} A signed graph $\mathrm{G}$ is {\it structurally  balanced} if there is  a partition of the node set into  $\mathrm{V}=\mathrm{V}_1\mcup \mathrm{V}_2$ with $\mathrm{V}_1$ and $\mathrm{V}_2$ being nonempty and  mutually disjoint,  where any edge between the two node subsets $\mathrm{V}_1$ and $\mathrm{V}_2$ is negative, and any edge within each $\mathrm{V}_i$ is positive.
\end{definition}

Known as the Harary's balance theorem, a signed graph $\mathrm{G}$ is structurally  balanced if and only if there is no cycle with an odd number of negative edges in $\mathrm{G}$ \cite{Cartwright-1956}. If $\mathrm{G}$ is a complete graph, it turned out that we can verify its structural balance property  by simply checking all triangles: $\mathrm{G}$ is structurally  balanced if and only if  among every set of three nodes there are either one or three positive edges \cite{David-2010}. The notion of structural balance can be weakened  in the following definition \cite{Davis-1967}.
\begin{definition} A signed graph $\mathrm{G}$ is {\it weakly structurally  balanced} if there is  a partition  $\mathrm{V}$ into  $\mathrm{V}=\mathrm{V}_1\mcup \mathrm{V}_2 \dots \mcup\mathrm{V}_m, m\geq 2$ with   $\mathrm{V}_1,\dots,\mathrm{V}_m$ being nonempty and  mutually disjoint,  where  any edge between different $\mathrm{V}_i$'s is negative, and any edge within each $\mathrm{V}_i$ is positive.
\end{definition}

\quad

\begin{figure}[h]
    \centering
    \begin{subfigure}[b]{0.3\textwidth}
        \includegraphics[width=\textwidth]{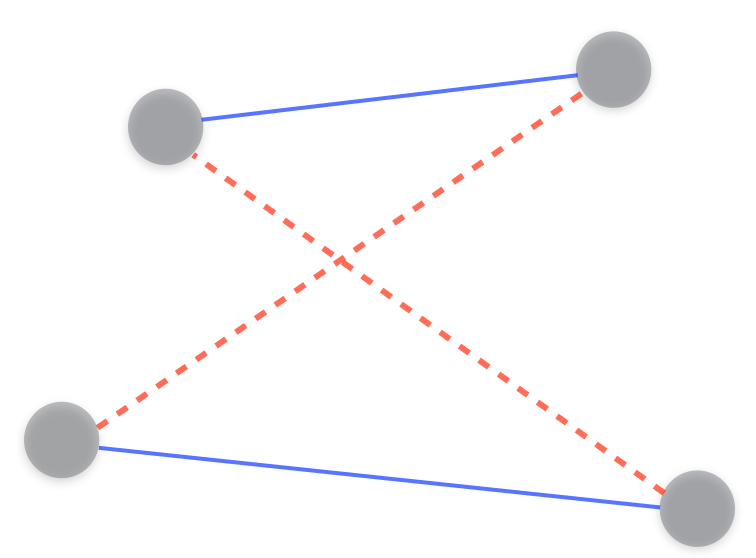}
        \caption{Strong balance.}
        \label{fig:gull}
    \end{subfigure}
    \begin{subfigure}[b]{0.3\textwidth}
        \includegraphics[width=\textwidth]{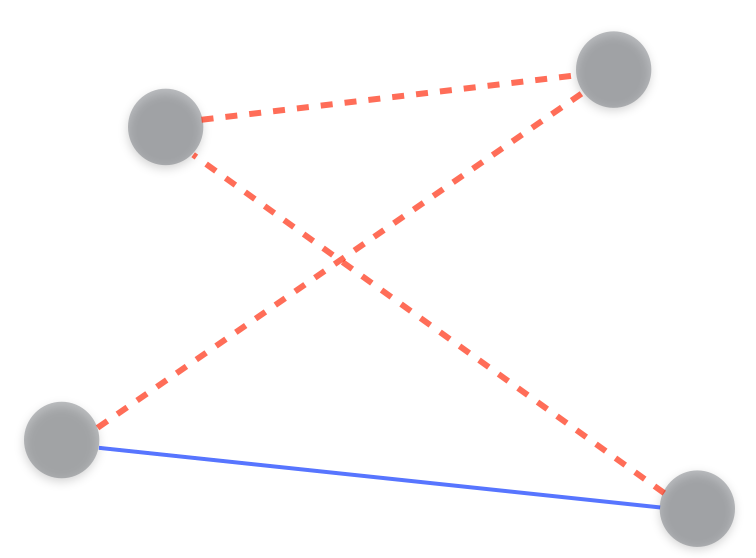}
        \caption{Weak balance.}
        \label{fig:tiger}
    \end{subfigure}
    \begin{subfigure}[b]{0.3\textwidth}
        \includegraphics[width=\textwidth]{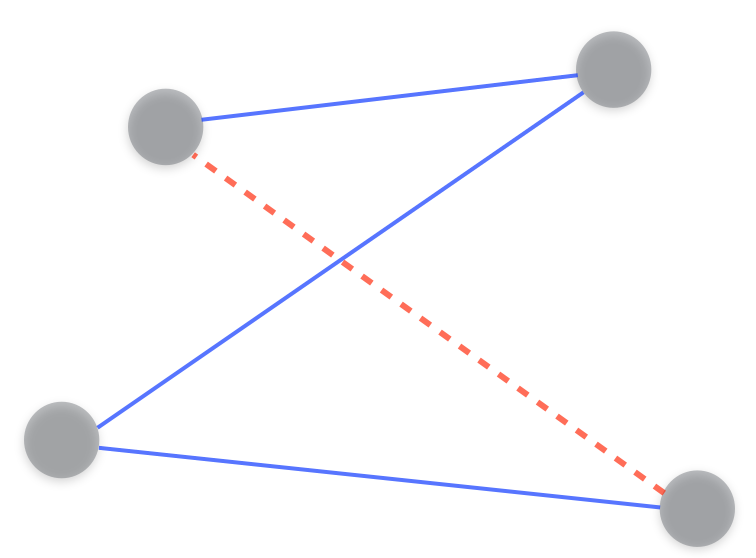}
        \caption{Unbalance.}
        \label{fig:mouse}
    \end{subfigure}
    \caption{Examples of strongly balanced (left), weakly balanced (middle), and unbalanced signed graphs (right). Here blue lines represent positive edges; red dashed lines represent negative edges. } \label{fig:graphs}
\end{figure}

It is known that  $\mathrm{G}$ is weakly structurally  balanced if and only if no cycle has exactly one negative edge in $\mathrm{G}$ \cite{Davis-1967}.  When $\mathrm{G}$ is a complete graph, this condition is equivalent to the fact that there is no set of three nodes among which there is exactly one negative edge \cite{David-2010}.
In Figure \ref{fig:graphs}, three basic  examples are presented illustrating graph   balance.

\subsection{Positive/Negative Interactions}
\label{sec:pos-neg-int}
Time is slotted at $t=0,1,\dots$. Each node $i$ holds a state $x_i(t)\in\Re$ at time $t$ and interacts with its neighbors at each time to revise its state. The interaction rule is specified by the sign of the links. Let $\alpha,\beta\geq 0$.
We first focus on a particular link $\{i,j\}\in\mathrm{E}$ and specify for the moment the dynamics along this link isolating all other interactions.
\begin{itemize}
\item  If the sign of $\{i,j\}$ is positive,   each node $s\in\{i,j\}$ updates its value by
\begin{itemize}
\item  The DeGroot Rule
\begin{align}\label{positive}
x_s(t+1)&=x_s(t) + \alpha \big(x_{-s}(t)-x_s(t)\big)=(1- \alpha)x_{s}(t)+ \alpha x_{-s}(t),
\end{align}
where ${-s}\in\{i,j\}\setminus \{s\}$ with $\alpha\in(0,1)$.
\end{itemize}

 \item  If the sign of $\{i,j\}$ is negative,   each node $s\in\{i,j\}$ updates its value by either
\begin{itemize}
\item The Opposing  Rule:
\begin{align}\label{s-f}
x_s(t+1)&=x_s(t)+\beta  \big(-x_{-s}(t)-x_s(t)\big)=(1-\beta)x_{s}(t)- \beta x_{-s}(t);
\end{align}
or
\item The Repelling  Rule:
\begin{align}\label{r-s-f}
x_s(t+1)&=x_s(t)- \beta  \big(x_{-s}(t)-x_s(t)\big)=(1+\beta)x_{s}(t)- \beta x_{-s}(t).
\end{align}
\end{itemize}

 \end{itemize}

 The positive interaction is consistent with  DeGroot's rule of social interactions, which indicates that the opinions of trustful social members are attractive to each other   \cite{Degroot_1974}. { Along a negative link, the opposing   rule (introduced in \cite{Altafini_TAC13} in the form of continuous-time dynamics) indicates that the interaction will drive a node state to be attracted by the opposite of its neighbor's state;  the repelling  rule  \cite{Shi-JSAC13-2} indicates that the two node states will  repel each other instead of being attractive.  The two parameters $\alpha$ and $\beta$ mark    the strength of positive and negative links, respectively. There can indeed be various types of negative interactions. As the DeGroot rule is the (discrete-time) gradient flow of the Laplacian quadratic form for networks  with only positive links \cite{Magnus-book}, the opposing rule and the repelling rule define network  gradient flows from the quadratic forms by the opposing and repelling Laplacians of signed graphs in (\ref{lq1}) and (\ref{lq2}), respectively. Therefore, these opposing/repelling rules are quite natural to be considered as the primary signed dynamic models, especially from the perspective of social opinion dynamics \cite{Altafini_TAC13,Shi-OR}.    }

\subsection{Paper Organization}
 This paper reviews the existing results on fundamental convergence  properties of signed dynamical networks \cite{Altafini_PONE12,Altafini_TAC13,Shi-JSAC13-2,Shi-TCNS15,Shi-TCNS17, Liu-Basar-15,Meng-SICON-2015,Meng-arXiv-2014,Shi-OR,Xia-Cao-TCNS-15,He14,Anderson-Shi-15}. In the past few years,  a variety of signed network models  appears in the literature that falls to the categories of the above opposing or repelling rules. Various treatments ranging from Lyapunov direct methods \cite{Altafini_TAC13} to graph  lifting \cite{He14} and even analysis based on complete observability theory \cite{Anderson-Shi-15} have been used to answer questions concerning with node state consensus  or clustering  in the asymptotic limit. We form a general signed network model by collecting  the node interactions at individual links of an underlying     graph. Then an algebraic-graphical method is provided  serving as a system-theoretic  tool for studying consensus  dynamics over signed networks. Combining generalized Perron-Frobenius theory, graph theory, and elementary algebraic recursions, we  show that this approach provides simple yet unified  proofs  to a series of basic convergence results for  networks with    deterministic or random node interactions.

The remainder of the paper is organized as follows. Section \ref{sec:deterministic} presents a series of basic results for dynamics over deterministic networks. Section \ref{sec:random} extends the discussion to random networks with  convergence results established using similar algebraic-graphical analysis  with a few additional probabilistic ingredients. Finally, Section \ref{sec:conclusions} concludes the paper with a few concluding  remarks in addition to some discussions on open problems and  future directions.

\subsection{Notation}
Real numbers are  in general  denoted by lowercase letters $x,y,a,b,c,...$ and  lowercase Greek letters $\alpha, \beta, \gamma,...$.  All vectors are column vectors denoted by bold lowercase letters $\mathbf{x},\mathbf{y},...$. Matrices   are denoted with upper case letters such as $A, B, C,...$. All matrices are real.  Given a matrix $A$, $A^\top$ denotes its transpose and $A^k$ denotes the $k$-th power of $A$ when it is a square matrix. Likewise the transpose of a vector $\mathbf{x}$ is denoted by $\mathbf{x}^\top$. The $ij$-entry of a matrix $A$ is denoted by $[A]_{ij}$; the spectrum and spectral radius of a matrix $A$ is denoted by $\sigma(A)$ and $\rho(A)$, respectively; the largest eigenvalue of a symmetric matrix $A$ is denoted by $\lambda_{\rm max}(A)$.   The $n$-dimensional all-one vector is denoted by $\mathbf{1}$, and the $n$-dimensional unit vector with the $i$'th entry being one is $\mathbf{e}_i$. The node set  is always $\mathrm{V}=\{1,\dots,n\}$, over which  a deterministic graph is  denoted as $\mathrm{G}$ and a random graph is denoted as $\mathpzc{G}$.   Depending on the argument, $|\cdot|$ stands for the absolute value of a real number or the cardinality of a set. The Euclidean norm of a vector is $\|\cdot\|$.

\section{Deterministic Networks}\label{sec:deterministic}
In this section, we investigate the evolution of the node states with deterministic interactions. The pairwise interactions among the signed links are collected over a deterministic network. We are interested in characterizing the asymptotic limits of the node states and provide some basic convergence theorems. Relevant  results in the literature can be seen for instance   in \cite{Altafini_TAC13, Liu-Basar-15,Meng-arXiv-2014,Xia-Cao-TCNS-15,He14}.

\subsection{Fundamental Convergence Results}
  \subsubsection{Opposing  Negative Dynamics}
With the opposing  rule (\ref{s-f}) along with the negative links, the update of $x_i(t)$ reads as
\begin{align}\label{s-f-deterministic}
x_i(t+1)&=x_i(t)+\alpha\sum_{j\in \mathrm{N}_i^+} \Big(x_j(t)-x_i(t) \Big)-\beta \sum_{j\in\mathrm{N}_i^-}\Big(x_j(t)+x_i(t)\Big)\nonumber\\
&=\Big(1-\alpha{\rm deg}_i^+-\beta {\rm deg}_i^-\Big)x_i(t)+\alpha \sum_{j\in \mathrm{N}_i^+}x_j(t)-\beta \sum_{j\in \mathrm{N}_i^-}x_j(t).
\end{align}
Denote $\mathbf{x}(t)=(x_1(t)\dots x_n(t))^\top$.  We can now rewrite (\ref{s-f-deterministic}) into the  following compact form:
\begin{align}\label{1}
\mathbf{x}(t+1)=W_{_\mathrm{G}}\mathbf{x}(t)=\big(I-\alpha{L}_{_{\mathrm{G}^+}}-\beta{L}_{_{\mathrm{G}^-}}^{^{\rm o}}\big)\mathbf{x}(t)
\end{align}{
where ${L}_{_{\mathrm{G}^+}}$ and ${L}_{_{\mathrm{G}^-}}^{^{\rm o}}$ are the opposing   Laplacians of $\mathrm{G}^+$ and $\mathrm{G}^-$, respectively. Also note that
$$
W_{_\mathrm{G}}=I-\alpha{L}_{_{\mathrm{G}^+}}-\beta{L}_{_{\mathrm{G}^-}}^{^{\rm o}}=I-{L}_{_{\mathrm{G}}}^{^{\rm ow}},
$$
with ${L}_{_{\mathrm{G}}}^{^{\rm ow}}=\alpha{L}_{_{\mathrm{G}^+}}+\beta{L}_{_{\mathrm{G}^-}}^{^{\rm o}}$ being the opposing weighted  Laplacian of $\mathrm{G}$. }

  Recall that a real matrix (or vector) is called positive (non-negative) if  all its entries are positive (non-negative); a stochastic matrix is a nonnegative matrix  with row sum equal to one \cite{Horn-1985}. A key property of the matrix  $W_{_\mathrm{G}}$ lies in that for small $\alpha$ and $\beta$, (e.g., $0<\alpha+\beta<{1}/\max_{i\in\mathrm{V}} {\rm deg}_i$)
\begin{align}\label{absolute-stochastic}
\sum_{j=1}^n \big|[W_{_\mathrm{G}}]_{ij}\big|=1,\ i\in\mathrm{V}
\end{align}
which indicates that $W_{_\mathrm{G}}$ will become a stochastic matrix if all its entries are put into their absolute values. { The following result holds relating the structural balance of $\mathrm{G}$ with the notion of bipartite consensus, i.e., node states are asymptotically clustered into two values with opposite signs. Such type of result was first presented in \cite{Altafini_TAC13} for continuous-time node dynamics based on Lyapunov analysis. Here we provide a proof by  incorporating graphical analysis into plain algebraic inequalities.
}

\begin{theorem}\label{theorem-sf-deterministic}
Assume that $0<\alpha+\beta<{1}/\max_{i\in\mathrm{V}} {\rm deg}_i$. Then along (\ref{s-f-deterministic}) the following statements hold for any initial value $\mathbf{x}(0)$.
\begin{itemize}
\item[(i)] If $\mathrm{G}$ is  structurally balanced subject to the partition $\mathrm{V}=\mathrm{V}_1\mcup \mathrm{V}_2$, then
$\lim_{t\to\infty} x_i(t)=\big(\sum_{j\in\mathrm{V}_1}x_j(0)-\sum_{j\in\mathrm{V}_2}x_j(0)\big)/n,\ i\in\mathrm{V}_1$,
 and $\lim_{t\to\infty} x_i(t)=-\big(\sum_{j\in\mathrm{V}_1}x_j(0)-\sum_{j\in\mathrm{V}_2}x_j(0)\big)/n, \ i\in\mathrm{V}_2$.
\item[(ii)] If $\mathrm{G}$ is not  structurally balanced, then $\lim_{t\to\infty} x_i(t)=0,\ i\in\mathrm{V}$.
\end{itemize}
\end{theorem}
{\it Proof.} \noindent (i) Let $\mathrm{G}$ be  structurally balanced with partition $\mathrm{V}=\mathrm{V}_1\mcup \mathrm{V}_2$. Consider a gauge transformation given by
$$
z_i(t)=x_i(t), i\in\mathrm{V}_1; \quad z_i(t)=-x_i(t), i\in\mathrm{V}_2.
$$
The evolution of the $z_i(t)$ becomes a  standard consensus algorithm, whose convergence follows  from for instance Theorem 2 in \cite{Olfati-Saber-2007}. The convergence of $x_i(t)$ can then be inferred.

\noindent (ii) Let $0<\alpha+\beta<{1}/{\rm deg}_i$ for all $i$. Applying Ger\v{s}hgorin's Circle Theorem (see, e.g., Theorem 6.1.1 in \cite{Horn-1985}), it is easy to see that $-1<\lambda_i(W_{_\mathrm{G}}) \leq 1$ for all
$\lambda_i\in \sigma(W_{_\mathrm{G}})$. This immediately implies that for any initial value $\mathbf{x}(0)$, there exists $\mathbf{y}(\mathbf{x}(0))=(y_1(\mathbf{x}(0))\ \dots \ y_n(\mathbf{x}(0)))^\top$ satisfying $W_{_\mathrm{G}}\mathbf{y}=\mathbf{y}$ such that
$
\lim_{t\to \infty} x_i(t)=y_i.
$

\noindent{\em Claim.} $|y_1|=\dots=|y_n|$ for any $\mathbf{x}(0)$.

Suppose there are two distinct nodes $i$ and $j$ with $|y_i|\neq |y_j|$. The fact that $W_{_\mathrm{G}}\mathbf{y}=\mathbf{y}$ gives
\begin{align}
|y_i|\leq \sum_{j=1}^n \big|[W_{_\mathrm{G}}]_{ij}\big|\cdot |y_j|,\ i\in\mathrm{V}.
\end{align}
This is impossible for a connected graph $\mathrm{G}$   noting (\ref{absolute-stochastic}). This proves the above claim.

Now let $y_\ast=|y_1|=\dots=|y_n|\neq 0$ for some $\mathbf{x}(0)$. There must be a set $\mathrm{V}_\ast$ (which, of course, may be an empty set at this point) with
$$
y_i= y_\ast, i\in \mathrm{V}_\ast; \ y_i= -y_\ast, i\in \mathrm{V}\setminus\mathrm{V}_\ast.
$$
It is straightforward to verify that in order for $W_{_\mathrm{G}}\mathbf{y}=\mathbf{y}$ to hold, all links (if any) in either $\mathrm{V}_\ast$ or $\mathrm{V}\setminus\mathrm{V}_\ast$ must be positive, and the links (if any) between $\mathrm{V}_\ast$ and $\mathrm{V}\setminus\mathrm{V}_\ast$ must be negative. This is to say, $\mathrm{G}$ must be  structurally balanced since by our standing assumption $\mathrm{G}^-$ is nonempty. We have now completed the proof. \hfill$\square$

We remark that the condition $0<\alpha+\beta<{1}/\max_{i\in\mathrm{V}} {\rm deg}_i$ in Theorem \ref{theorem-sf-deterministic} can be certainly relaxed, e.g., a straightforward one would be $0<\alpha{\rm deg}_i^++\beta {\rm deg}_i^-<1$ for all $i$. Further relaxations can be obtained making use of  the structure of ${L}_{_{\mathrm{G}^+}}$ and ${L}_{_{\mathrm{G}^-}}^{^{\rm o}}$, and  the fact that  the spectrum of $W_{_\mathrm{G}}$ will be restricted within the unit cycle for sufficiently small $\alpha$ and $\beta$. The essential message of Theorem \ref{theorem-sf-deterministic} is that structural balance of $\mathrm{G}$ determines whether one is within the spectrum of $W_{_\mathrm{G}}$. In fact, there holds
\begin{align}\label{sf-contraction}
\big\|\mathbf{x}(t+1)\big\|^2 \leq \lambda_{\rm max}\big(W_{_\mathrm{G}}^2\big)\big\|\mathbf{x}(t)\big\|^2 \leq \big\|\mathbf{x}(t)\big\|^2
\end{align}
with sufficiently small $\alpha$ and $\beta$ guaranteeing $\lambda_{\rm max}\big(W_{_\mathrm{G}}^2\big)\leq 1$. Therefore,  the algorithm (\ref{1}) defines an overall contraction mapping, consistent with the standard consensus algorithms without negative links.

\subsubsection{Repelling  Negative Dynamics}

Now consider the repelling  rule (\ref{r-s-f})  for negative links. The update of $x_i(t)$ reads as
\begin{align}\label{r-s-f-deterministic}
x_i(t+1)&=x_i(t)+\alpha\sum_{j\in \mathrm{N}_i^+} \Big(x_j(t)-x_i(t) \Big)-\beta \sum_{j\in\mathrm{N}_i^-}\Big(x_j(t)-x_i(t)\Big)\nonumber\\
&=\Big(1-\alpha{\rm deg}_i^++\beta {\rm deg}_i^-\Big)x_i(t)+\alpha \sum_{j\in \mathrm{N}_i^+}x_j(t)-\beta \sum_{j\in \mathrm{N}_i^-}x_j(t).
\end{align}
The algorithm (\ref{r-s-f-deterministic}) can be  written into
\begin{align}\label{2}
\mathbf{x}(t+1)=M_{_\mathrm{G}}\mathbf{x}(t)=\big(I-\alpha{L}_{_{\mathrm{G}^+}}-\beta{L}_{_{\mathrm{G}^-}}^{^{\rm r}}\big)\mathbf{x}(t).
\end{align}
Here $$M_{_\mathrm{G}}=I-\alpha{L}_{_{\mathrm{G}^+}}-\beta{L}_{_{\mathrm{G}^-}}^{^{\rm r}}=I-{L}_{_{\mathrm{G}}}^{^{\rm rw}},
$$
with ${L}_{_{\mathrm{G}}}^{^{\rm rw}}=\alpha{L}_{_{\mathrm{G}^+}}+\beta{L}_{_{\mathrm{G}^-}}^{^{\rm r}}$ being the repelling  weighted  Laplacian of $\mathrm{G}$. From (\ref{2}),  $M_{_\mathrm{G}}\mathbf{1}=\mathbf{1}$ always holds. { We present the following result, which  by itself is merely a straightforward look into the spectrum of the repelling Laplacian ${L}_{_{\mathrm{G}}}^{^{\rm rw}}$. }

\begin{theorem}\label{theorem-rsf-deterministic}
  Suppose $\mathrm{G}^+$ is connected. Then along (\ref{r-s-f-deterministic}) for any  $0<\alpha<1/\max_{i\in\mathrm{V}} {\rm deg}_i^+$, there exists a critical value $\beta_\ast>0$ for $\beta$ such that
   \begin{itemize}
   \item[(i)] If $\beta<\beta_\ast$, then  average consensus is reached  in the sense that
  $
  \lim_{t\to \infty} x_i(t)={\sum_{j=1}^n x_i(0)}/{n}
  $
 for all initial values $\mathbf{x}(0)$;
 \item[(ii)]   If $\beta>\beta_\ast$, then $\lim_{t\to\infty}\big\|\mathbf{x}(t)\big\|=\infty$ for almost all initial values w.r.t. Lebesgue  measure.
 \end{itemize}
\end{theorem}
{\it Proof.}  Define $J=\mathbf{1}\mathbf{1}^\top/n$. Fix $\alpha\in(0,1/\max_{i\in\mathrm{V}} {\rm deg}_i^+)$ and consider
$$
f(\beta):=\lambda_{\rm max}\Big(I-\alpha{L}_{_{\mathrm{G}^+}}-\beta{L}_{_{\mathrm{G}^-}}^{^{\rm r}}-J\Big), \ \ g(\beta):=\lambda_{\rm min}\Big(I-\alpha{L}_{_{\mathrm{G}^+}}-\beta{L}_{_{\mathrm{G}^-}}^{^{\rm r}}-J\Big).
$$
The Courant-Fischer Theorem (see Theorem 4.2.11 in \cite{Horn-1985}) implies
 that both $f(\cdot)$ and $g(\cdot)$ are  continuous and non-decreasing functions over $[0,\infty)$. The matrix $J$ always commutes with 
 $I-\alpha{L}_{_{\mathrm{G}^+}}-\beta{L}_{_{\mathrm{G}^-}}^{^{\rm r}}$, and  $1$ is the only nonzero eigenvalue of $J$. Moreover, the eigenvalue $1$ of $J$ shares a common eigenvector $\mathbf{1}$ with 
 the eigenvalue $1$ of $I-\alpha{L}_{_{\mathrm{G}^+}}-\beta{L}_{_{\mathrm{G}^-}}^{^{\rm r}}$.  
 
 Since $\mathrm{G}^+$ is connected,  the second smallest eigenvalue of ${L}_{_{\mathrm{G}^+}}$ is positive.   Since  $0<\alpha<1/\max_{i\in\mathrm{V}} {\rm deg}_i^+$, there holds $\lambda_{\rm min}\big( I-\alpha{L}_{_{\mathrm{G}^+}}\big)>-1$   again due to  the Ger\v{s}hgorin's Circle Theorem. Therefore, $f(0)<1$, $g(0)>-1$. 
Noticing  $f(\infty)=\infty>1$, there exists   $\beta_\ast>0$ satisfying  $f(\beta_\ast)=1$. We can then verify  the following facts. 
\begin{itemize}
\item There hold $f(\beta) <1$ and $g(\beta)>-1$ if $\beta<\beta_\ast$. In this case, along (\ref{2}) $\lim_{t\to\infty}(I-J)\mathbf{x}(t)=0$, which in turn implies that $\mathbf{x}(t)$ converges to the eigenspace corresponding to the eigenvalue one of  $M_{_\mathrm{G}}$. This  leads to the average consensus statement in (i).

\item There holds $f(\beta) >1$ if $\beta>\beta_\ast$. In this case, along (\ref{2})  $\mathbf{x}(t)$ diverges as long as the initial value $\mathbf{x}(0)$ has a nonzero projection onto the eigenspace corresponding to $\lambda_{\rm max}\big(M_{_\mathrm{G}}\big)$ of $M_{_\mathrm{G}}$. This leads to the almost everywhere divergence statement in (ii).
\end{itemize}
The proof is now complete. \hfill$\square$

The condition that $\mathrm{G}^+$ is a connected graph is crucial for Theorem \ref{theorem-rsf-deterministic}.  Once $\mathrm{G}^+$ becomes disconnected, it is easy to see that one single negative link and an arbitrarily small $\beta>0$ will drive the network state to diverge  for almost all initial values. { Necessary and sufficient conditions are established in \cite{chen-cdc-2016,zelazo2017} on when the repelling Laplacian ${L}_{_{\mathrm{G}}}^{^{\rm rw}}$ is positive semidefinite from linear matrix inequalities, which can be utilized to establish deeper results compared to Theorem \ref{theorem-rsf-deterministic}. Also see \cite{siam-2014-spectrum} for a much more detailed    spectrum analysis of repelling Laplacians.
}

\subsection{Mathematical Reasoning: Eventually Positive Matrices}
Theorems \ref{theorem-sf-deterministic} and \ref{theorem-rsf-deterministic} provide some basic yet informative characterizations to how negative links influence the network dynamics in the two  models:
\begin{itemize}
\item With opposing   rule, both the positive and negative links contribute to state convergence of the nodes. The overall dynamics has a contraction nature with small $\alpha$ and $\beta$. As long as the overall graph $\mathrm{G}$ is connected, the absolute values of  node states asymptotically agree; structural balance of the graph merely further determines the existence of nontrivial absolute value agreement in the sense that a bipartite consensus is achieved.

\item With repelling   negative dynamics,  the negative links produce repulsive interactions with a divergence nature. These negative links are therefore essentially  perturbations: the positive links must generate convergence with sufficient speed so that the negative links can be overcome. This requires that the positive graph $\mathrm{G}^+$ must be connected by itself and  results in the critical value of $\beta$ below which convergence to consensus still holds.
\end{itemize}

It has been well known that convergence of standard consensus algorithms is closely related to the Perron--Fronenius Theory \cite{Olfati-Saber-2007}. Consider a graph $\mathrm{G}$ (unsigned) with  Laplacian $L_{_\mathrm{G}}$.  A standard consensus algorithm over the graph $\mathrm{G}$, is defined as
 \begin{align}\label{3}
 x_i(t+1)=x_i(t)+\alpha \sum_{j\in \mathrm{N}_i}\big(x_j(t)-x_i(t)\big),\ i\in\mathrm{V}
 \end{align}
or in vector form,
 \begin{align}
 \mathbf{x}(t+1)=S_{_\mathrm{G}} \mathbf{x}(t),
 \end{align}
 where $S_{_\mathrm{G}}=I- \alpha L_{_\mathrm{G}}$. Obviously,  $S_{_\mathrm{G}}$  is a non-negative matrix for $\alpha<1/\max_{i\in\mathrm{V}}{\rm deg}_i$. 
 The  Perron--Fronenius Theory is the   fundamental reasoning behind  the convergence  of the algorithm (\ref{3}) \cite{Olfati-Saber-2007}: if  and only if  $\mathrm{G}$ is connected, there holds $$
 \lim_{t\to \infty} S_{_\mathrm{G}} ^t =\mathbf{1}\mathbf{1}^\top/n.
  $$
  In fact, $\mathbf{1}^\top$ and $\mathbf{1}$ are the left and right eigenvector  corresponding to eigenvalue $1$ of $S_{_\mathrm{G}}$, known as  its Perron--Frobenius eigenvalue.

A matrix $A$ is called eventually positive  if there exists an integer  $k_0 \in\mathbb{N}^+$ such that $A^k$  is positive for all $k\geq k_0$. If $\mathrm{G}$ is structurally balanced subject to node set partition $\mathrm{V}_1$ and $\mathrm{V}_2$, it is easy to see that $KW_{_\mathrm{G}}K^{-1}$ defines a nonnegative  stochastic matrix,  which is  eventually positive  if $0<\alpha+\beta<{1}/\max_{i\in\mathrm{V}} {\rm deg}_i$, where $K={\rm diag}(k_1,\dots,k_n)$ with $k_i=1,i\in\mathrm{V}_1$ and $k_i=-1,i\in\mathrm{V}_2$.
 On the other hand, the matrix $M_{_\mathrm{G}}$ for repelling rule would contain  negative values. Letting  $\beta_\ast$ be the critical value established in Theorem \ref{theorem-rsf-deterministic}, the following conclusion shows that $M_{_\mathrm{G}}$ is also  eventually positive when convergence is achieved. We refer to \cite{Altafini_TAC15} for a deeper  investigation on eventual positiveness of signed network dynamics.


\begin{proposition}\label{proposition-eventually-positive} Let $\mathrm{G}^+$ be connected. Then  $M_{_\mathrm{G}}=I-\alpha{L}_{_{\mathrm{G}^+}}-\beta{L}_{_{\mathrm{G}^-}}^{^{\rm r}}$ is eventually positive   if  $0<\alpha<1/\max_{i\in\mathrm{V}} {\rm deg}_i$ and $\beta<\beta_\ast$.
\end{proposition}{
{\it Proof.} Note that (see Theorem 2.2 in \cite{Noutsos-2006}), a matrix $A\in\mathbb{R}^{n\times n}$ is eventually positive if  both $A$ and $A^\top$ have the strong Perron--Frobenius property: (i) $\rho(A)$ is a simple positive eigenvalue of $A$; (ii) the right eigenvector related to $\rho(A)$ is positive. The  statement is immediate by verifying that  $M_{_\mathrm{G}}$ has the Perron--Frobenius property under the given conditions, respectively, from the proof of Theorem  \ref{theorem-rsf-deterministic}.} \hfill$\square$

\subsection{Directed Graphs}
Directional links in a network can also be associated with signs \cite{Wasserman-book-1994}. We now present generalizations of the previous model and results to signed directed networks. For the ease of presentation, we keep the previous notation and simply adapt them to the directed graph case. Their usage is of course restricted to the current subsection.

Now let the graph $\mathrm{G}=(\mathrm{V},\mathrm{E})$ be a directed graph (digraph), where a link $(i,j)\in\mathrm{E}$ is directed starting from $i$ and pointing to $j$. A diagraph is termed a signed digraph if each of its link has a positive or negative sign. By revising the definition of positive and negative neighbor sets of node $i$ to
 $$
\mathrm{N}_i^+:=\big\{j:(j,i)\in\mathrm{E}^+\big\}; \quad \mathrm{N}_i^-:=\big\{j:(j,i)\in\mathrm{E}^-\big\},
$$
the network dynamics (\ref{s-f-deterministic}) and (\ref{r-s-f-deterministic}) are then readily  defined for the digraph $\mathrm{G}$. The set  $\mathrm{N}_i=\mathrm{N}_i^+\mcup \mathrm{N}_i^-$ continues to represent the overall neighbor  set of node $i$. In this directed graph case we continue to define ${\rm deg}_i^+=|\mathrm{N}_i^+|$, ${\rm deg}_i^-=|\mathrm{N}_i^-|$, and   ${\rm deg}_i=|\mathrm{N}_i|$ as the positive, negative, and overall degrees of node $i$. We can also define the degree matrices ${D}_{_{\mathrm{G}^+}}$ and ${D}_{_{\mathrm{G}^-}}$ based on these positive or negative degrees.

The concept of structural balance  can  be generalized to digraphs by replacing the undirected edges with directional links.

\begin{definition}A signed digraph $\mathrm{G}$ is {\it structurally  balanced} if there is  a partition of the node set into  $\mathrm{V}=\mathrm{V}_1\mcup \mathrm{V}_2$ with $\mathrm{V}_1$ and $\mathrm{V}_2$ being nonempty and    disjoint,  such that any directional link between  $\mathrm{V}_1$ and $\mathrm{V}_2$ is negative, and any link with two end nodes belonging to the same $\mathrm{V}_i$ is positive.
\end{definition}

{
For a digraph $\mathrm{G}$, the adjacency matrix ${A}_{_{\mathrm{G}^+}}$ of $\mathrm{G}^+$ is given by $[{A}_{_{\mathrm{G}^+}}]_{ij}=1$ if $(j,i)\in\mathrm{E}^+$ and $[{A}_{_{\mathrm{G}^+}}]_{ij}=0$ otherwise;  the   adjacency matrix ${A}_{_{\mathrm{G}^-}} $ of $\mathrm{G}^-$ is given by $[{A}_{_{\mathrm{G}^-}}]_{ij}=-1$ if $(j,i)\in\mathrm{E}^-$ and $[{A}_{_{\mathrm{G}^+}}]_{ij}=0$ otherwise.  Then $
{L}_{_{\mathrm{G}^+}}:={D}_{_{\mathrm{G}^+}}-{A}_{_{\mathrm{G}^+}}
$ is the Laplacian of  the  directed positive subgraph, and  $$
{L}_{_{\mathrm{G}^-}}^{^{\rm o}}:={D}_{_{\mathrm{G}^-}}-{A}_{_{\mathrm{G}^-}}
$$
is the opposing Laplacian of the   directed negative subgraph. The dynamics (\ref{s-f-deterministic}) can still be written  into the form of (\ref{1}) with $W_{_\mathrm{G}}=I-\alpha{L}_{_{\mathrm{G}^+}}-\beta{L}_{_{\mathrm{G}^-}}^{^{\rm o}}$.  The following theorem is a generalization to Theorem \ref{theorem-sf-deterministic} for signed digraphs.
}
\begin{theorem}\label{theorem-sf-deterministic-directed}
Consider network dynamics (\ref{s-f-deterministic}) over a digraph $\mathrm{G}$. Assume that $0<\alpha+\beta<{1}/\max_{i\in\mathrm{V}} {\rm deg}_i$. Suppose $\mathrm{G}$ is strongly connected. The following statements hold for any initial value $\mathbf{x}(0)$.
\begin{itemize}
\item[(i)] If $\mathrm{G}$ is  structurally balanced subject to partition $\mathrm{V}=\mathrm{V}_1\mcup \mathrm{V}_2$, then there are $n$ positive numbers $w_1,\dots,w_n$ with $\sum_{i=1}^nw_i=1$ such that $\lim_{t\to\infty} x_i(t)=\big(\sum_{j\in\mathrm{V}_1} w_jx_j(0)-\sum_{j\in\mathrm{V}_2}w_jx_j(0)\big)/n$, $i\in\mathrm{V}_1$ and $\lim_{t\to\infty} x_i(t)=-\big(\sum_{j\in\mathrm{V}_1}w_jx_j(0)-\sum_{j\in\mathrm{V}_2}w_jx_j(0)\big)/n$,  $i\in\mathrm{V}_2$.
\item[(ii)] If $\mathrm{G}$ is not  structurally balanced, then $\lim_{t\to\infty} x_i(t)=0,\ i\in\mathrm{V}$.
\end{itemize}
\end{theorem}

The $(w_1 \dots w_n)$ in Theorem \ref{theorem-sf-deterministic-directed} is the left eigenvector related to the eigenvalue $1$ of  the matrix $KW_{_\mathrm{G}}K^{-1}$, which of course  depends on $\alpha$ and $\beta$. Again Ger\v{s}hgorin's Circle Theorem leads to   $\rho\big(W_{_\mathrm{G}}\big)\leq 1$. However, the matrix $W_{\rm G}$ with a directed graph $\mathrm{G}$ is no longer necessarily symmetric. We cannot immediately conclude from $\rho\big(W_{_\mathrm{G}}\big)\leq 1$ the state-convergence of the nodes as  in the proof of Theorem \ref{theorem-sf-deterministic} for undirected graphs. We can however bypass this obstacle by imposing a contradiction argument again from an algebraic-graphical recursion.

With a diagraph $\mathrm{G}^-$,   $$
{L}_{_{\mathrm{G}^-}}^{^{\rm r}}=-{D}_{_{\mathrm{G}^-}}-{A}_{_{\mathrm{G}^-}}
 $$
 is its repelling   Laplacian. The network dynamics (\ref{r-s-f-deterministic}) can be again represented by (\ref{2}) with $$
 M_{_\mathrm{G}}=I-\alpha{L}_{_{\mathrm{G}^+}}-\beta{L}_{_{\mathrm{G}^-}}^{^{\rm r}}.
   $$
   With $\mathrm{G}$ being directed, $M_{_\mathrm{G}}$ is not necessarily symmetric,  $M_{_\mathrm{G}} \mathbf{1} =\mathbf{1}$ however continues to hold. The following theorem corresponds to Theorem \ref{theorem-rsf-deterministic} for signed digraphs.
\begin{theorem}\label{theorem-rsf-deterministic-directed}
 Consider network dynamics (\ref{r-s-f-deterministic}) over a digraph $\mathrm{G}$. Suppose $\mathrm{G}^+$ is strongly connected and fix $0<\alpha<1/\max_{i\in\mathrm{V}} {\rm deg}_i^+$. There exists  $\beta_\ast>0$   such that for any $\beta<\beta_\ast$,  there are $q_1(\beta),\dots, q_n(\beta)\in\mathbb{R}^+$ with $\sum_{i=1}^n q_i(\beta)=1$  for which a consensus is reached at
  $$
  \lim_{t\to \infty} x_i(t)={\sum_{j=1}^n q_i(\beta)x_i(0)},  \ i\in\mathrm{V}
  $$
 for all initial values $\mathbf{x}(0)$.
\end{theorem}

In the statement of Theorem \ref{theorem-rsf-deterministic-directed}, for any $\beta<\beta_\ast$, $\big(q_1(\beta) \dots q_n(\beta))$ is a left eigenvector related to eigenvalue $1$ of $M_{_\mathrm{G}}$.
 It is worth emphasizing that the $\beta_\ast$ in Theorem \ref{theorem-rsf-deterministic-directed} is merely an upper bound for $\beta$ under which the network can still reach a consensus in the presence of the negative links, and it is unclear whether  such $\beta_\ast$ would remain   a critical value as the undirected case. {From the proof, the actual value of $\beta_\ast$ can be expressed by
$$
\sup_{\eta}\Big\{\eta:\ \max_{\lambda \in \sigma(M_{_\mathrm{G}})\setminus\{1\}}\big|\lambda\big|<1\ \mbox{for all}\ \beta<\eta\Big\}.
$$

 Theorem \ref{theorem-sf-deterministic-directed} is a  special case of various results in the literature \cite{Meng-arXiv-2014,Xia-Cao-TCNS-15,He14}, for which the same algebraic-graphical analysis can be adopted. Theorem \ref{theorem-rsf-deterministic-directed} follows from a straightforward matrix perturbation analysis.  The proofs of   Theorem \ref{theorem-sf-deterministic-directed} and Theorem \ref{theorem-rsf-deterministic-directed} have been put in the Appendix.
}

{
\subsection{Rates of Convergence}
The convergence statements throughout Theorem \ref{theorem-sf-deterministic} -- Theorem \ref{theorem-rsf-deterministic-directed} are of course exponential since the network dynamics are linear time-invariant. In either undirected or directed case, the rate of convergence of the network dynamics (whenever convergence has been assured) is specified by
\begin{itemize}
\item
$
\rho(W_{_\mathrm{G}})
$ under the opposing rule without structural balance;
\item
$
\rho(KW_{_\mathrm{G}}K^{-1}-\mathbf{1}\mathbf{1}^\top/n)
$ under the opposing rule with structural balance, where $K$ is the corresponding Gauge transform;
\item
$
\rho(M_{_\mathrm{G}} -\mathbf{1}\mathbf{1}^\top/n)
$ under the repelling rule.
\end{itemize}
From the structure of  $W_{_\mathrm{G}}$ and $M_{_\mathrm{G}}$, one can infer that for small $\alpha,\beta$ and with undirected node interactions, adding one link (positive or negative) for the opposing negative dynamics with structural balance will accelerate the convergence if structural balance is preserved;
adding one negative link for the repelling rule will always slow down convergence. The interplay between the weights $\alpha$ and $\beta$ and the positioning of the positive and negative links is however rather complex, which relies on how much the   spectrum analysis of repelling Laplacian  as in \cite{chen-cdc-2016,zelazo2017,siam-2014-spectrum} can be push forward. }
\subsection{Weighted Signs, Continuous-time Dynamics, Switching Structures}

More sophisticated signed networks can certainly be studied using similar  tools and analysis. This subsection presents a coverage to related  results in the literature.

\subsubsection{Weighted Signs}\label{Subsection:weighted}
The strength of positive and negative links, represented by $\alpha$ and $\beta$, can also be link dependent. This means that for the positive and negative dynamics (\ref{positive}), (\ref{s-f}), and (\ref{r-s-f}) along the edge $\{i,j\}$, $\alpha$ and $\beta$ will be replaced by  $\alpha_{ij}$ and $\beta_{ij}$, respectively. The results of Theorems \ref{theorem-sf-deterministic}--\ref{theorem-rsf-deterministic-directed} can be extended to networks with weighted signs straightforwardly \cite{Altafini_TAC13}.

\subsubsection{Continuous-time Dynamics}
The signed network dynamics considered above clearly have their continuous-time counter part. For the opposing  negative dynamics (\ref{1}), the corresponding  node state evolution in continuous time reads as
\begin{align}\label{c-1}
\frac{d}{dt}{\mathbf{x}(t)}=-\Big(\alpha{L}_{_{\mathrm{G}^+}}+\beta{L}_{_{\mathrm{G}^-}}^{^{\rm o}}\Big)\mathbf{x}(t)= -{L}_{_{\mathrm{G}}}^{^{\rm ow}}\mathbf{x}(t).
\end{align}
On the other hand, the continuous-time counter part of the repelling dynamics (\ref{2}) is
\begin{align}\label{c-2}
\frac{d}{dt}{\mathbf{x}(t)}=-\Big(\alpha{L}_{_{\mathrm{G}^+}}+\beta{L}_{_{\mathrm{G}^-}}^{^{\rm r}}\Big)\mathbf{x}(t)=-{L}_{_{\mathrm{G}}}^{^{\rm rw}}\mathbf{x}(t).
\end{align}

Evidently, the asymptotic behavior of (\ref{c-1}) and (\ref{c-2}) is fully determined by the spectrum of the Laplacian ${L}_{_{\mathrm{G}}}^{^{\rm ow}}$ and the repelling Laplacian ${L}_{_{\mathrm{G}}}^{^{\rm rw}}$. They are in fact shifts of  the spectrum of  $W_{_{\mathrm{G}}}$ and $M_{_{\mathrm{G}}}$, respectively. With continuous-time dynamics, we no longer need to worry about that  certain eigenvalues     be outside the unit cycle for large $\alpha$ and $\beta$. Consequently,  Theorems \ref{theorem-sf-deterministic} and \ref{theorem-rsf-deterministic} can be immediately  translated to the following statements.

\begin{proposition}
(i) Along the continuous-time evolution  (\ref{c-1}), the following hold for any initial value $\mathbf{x}(0)$:
\begin{itemize}
\item If $\mathrm{G}$ is  structurally balanced subject to partition $\mathrm{V}=\mathrm{V}_1\mcup \mathrm{V}_2$, then
$\lim_{t\to\infty} x_i(t)=\big(\sum_{j\in\mathrm{V}_1}x_j(0)-\sum_{j\in\mathrm{V}_2}x_j(0)\big)/n,\ i\in\mathrm{V}_1$
 and
 $\lim_{t\to\infty} x_i(t)=-\big(\sum_{j\in\mathrm{V}_1}x_j(0)-\sum_{j\in\mathrm{V}_2}x_j(0)\big)/n, \ i\in\mathrm{V}_2$.
\item If $\mathrm{G}$ is not  structurally balanced, then $
\lim_{t\to\infty} x_i(t)=0,\ i\in\mathrm{V}$.
\end{itemize}

\noindent (ii)  Consider  (\ref{c-2}) and suppose $\mathrm{G}^+$ is connected. Then for any  $\alpha>0$, there exists a critical value $\beta_\ast>0$ for $\beta$ such that
   \begin{itemize}
   \item If $\beta<\beta_\ast$, then an average consensus is reached, i.e.,
  $
  \lim_{t\to \infty} x_i(t)={\sum_{j=1}^n x_i(0)}/{n}$
 for all initial value $\mathbf{x}(0)$.
 \item  If $\beta>\beta_\ast$, then $\lim_{t\to\infty}\big\|\mathbf{x}(t)\big\|=\infty$ for almost all initial values w.r.t. Lebesgue measure.
 \end{itemize}
\end{proposition}

The results for opposing  negative dynamics can even be extended to nonlinear node interactions \cite{Altafini_PONE12,Meng-SICON-2015}. As illustrated in (\ref{sf-contraction}), under the opposing  negative dynamics, both positive and negative links lead to non-expansive  network state evolution\footnote{With directed graphs,  (\ref{sf-contraction}) in general no longer holds under the opposing  negative dynamics. However, there still holds that $\max_{i\in\mathrm{V}} \big|x_i(t+1)\big|\leq \max_{i\in\mathrm{V}} \big|x_i(t)\big|$ as shown in the proof of Theorem \ref{theorem-sf-deterministic-directed}. Therefore, the network  state evolution continues to be non-expansive.}. The mathematical reasoning  behind those non-linear generalizations is due to  the fact that the non-expansive property can be  preserved for suitable nonlinear interaction rules.

\subsubsection{Switching Network Structures}
In the study of standard consensus algorithms, one particular interest was to establish convergence conditions under time-varying  network structures \cite{Jadbabaie2003,Blondel-CDC-2005,Ren2005,Moreau2005}, for which earlier work was dated to 1960s \cite{Wolfowitz-1963}. Such analysis can be  challenging due to the absence of  a common convergence metric that works for all possible choices of   interaction graphs. Nevertheless, possibilities of generalizing the analysis of time-varying network structures have been shown in the literature \cite{Altafini_TAC13,Cao-TAC-15,Meng-SICON-2015,Xia-Cao-TCNS-15,Liu-Basar-15,Anderson-Shi-15}.

Let $\mathrm{G}_t=(\mathrm{V}, \mathrm{E}_t),t=0,1,\dots$ be a sequence of graphs with each $\mathrm{G}_t$ being a (directed or undirected) signed graph. Then the positive and negative neighbor sets of node $i$, are determined by connections in $\mathrm{G}_t$ and therefore become time-dependent, denoted $\mathrm{N}_i^+(t)$ and $\mathrm{N}_i^-(t)$, respectively.  The network dynamics under the opposing  rule (\ref{s-f}) are then represented by
 \begin{align}\label{s-f-deterministic-switching}
x_i(t+1)&=x_i(t)+\alpha\sum_{j\in \mathrm{N}_i^+(t)} \Big(x_j(t)-x_i(t) \Big)-\beta \sum_{j\in\mathrm{N}_i^-(t)}\Big(x_j(t)+x_i(t)\Big).
\end{align}
We cite the following result from  Theorem 2.1 and Theorem 2.2 in \cite{Meng-arXiv-2014}.

\begin{proposition}\label{proposition-switching-s-f}Suppose there exists a constant $0<\delta<1$ such that $\alpha \big| \mathrm{N}_i^+(t)\big|+ \beta \big| \mathrm{N}_i^-(t)\big|\leq 1-\delta$ for all $i\in\mathrm{V}$ and all $t\geq 0$.

(i)
 Let there exist $T\geq 0$ such that the graph $\mathrm{G}_{[s,s+T]}:=\big(\mathrm{V}, \mcup_{t=s}^{s+T} \mathrm{E}_t\big)$ is strongly connected for all $s\geq 0$. Then along (\ref{s-f-deterministic-switching}), for any initial value $\mathbf{x}(0)$, there exists $y_\ast(\mathbf{x}(0)) \geq 0$ such that $\lim_{t\to \infty} \big|x_i(t)\big|=y_\ast(\mathbf{x}(0))$ for all $i\in\mathrm{V}$.

(ii) Suppose $\mathrm{G}_t$ is undirected for all $t\geq 0$. Let the graph $\mathrm{G}_{[s,\infty]}:=\big(\mathrm{V}, \mcup_{t=s}^{\infty} \mathrm{E}_t\big)$ be connected for all $s\geq 0$. Then along (\ref{s-f-deterministic-switching}), for any initial value $\mathbf{x}(0)$, there exists $y_\ast(\mathbf{x}(0)) \geq 0$ such that $\lim_{t\to \infty} \big|x_i(t)\big|=y_\ast(\mathbf{x}(0))$ for all $i\in\mathrm{V}$.
\end{proposition}

The structural balance condition can be generalized    to the sequence of graphs $\mathrm{G}_t=(\mathrm{V}, \mathrm{E}_t)$, under which  bipartite consensus result can be similarly established for opposing  negative dynamics \cite{Cao-TAC-15,Xia-Cao-TCNS-15,Liu-Basar-15}. On the other hand, for repelling  negative dynamics, analysis for switching network structures can be extremely challenging since the network state is no longer non-expansive in the presence of one single negative link. It turns out that in order to preserve convergence to consensus, it is important that at each time step, the influence of the negative links can be overcome by the positive links. We refer to \cite{Anderson-Shi-15,ortega-tac-2018} for such treatment of continuous-time node dynamics.

\section{Random Networks}\label{sec:random}

Node interactions happen randomly  in many real-world networks, and how consensus can be reached over a random node interaction process has been extensively studied \cite{Hatano-TAC-2005,Boyd-TIT-2006,Fagnani-JSAC-2008,Tahbaz-Salehi-TAC-2008,Tahbaz-Salehi-TAC-2010,Kar-TSP-2010,Shi-IT}. We proceed to discuss network dynamics over signed random graph processes, where relevant    results  appeared  in  \cite{Shi-JSAC13-2,Shi-TCNS15,Shi-TCNS17, Liu-Basar-15,Shi-OR}.

We use the following gossiping model \cite{Boyd-TIT-2006} to describe the random node interactions. The undirected, signed graph, $\mathrm{G}=(\mathrm{V}, \mathrm{E})$, continue to define the world of the network where interactions take place. Each node initiates interactions at the instants of a rate-one Poisson process, and at each of these instants, picks a node at random to interact with. Under this model, at a given time, at most one node initiates an interaction. This allows us to order interaction events in time and to focus on modeling the node pair selection at the interaction times. The node pair selection is then characterized as follows.

\begin{definition} Independently at each interaction event $t\geq0$, (i) a node $i\in\mathrm{V}$ is drawn uniformly at random, i.e., with probability $1/n$; (ii) node $i$ picks a neighbor $j$ uniformly with probability $1/{\rm deg}_i$ for $j\in \mathrm{N}_i$. In this case, we say that the unordered node pair $\{i,j\}$ is selected.
\end{definition}

Let $(\mathrm{E}, \mathscr{S}, \mu)$ be the probability space, where $\mathscr{S}$ is the discrete $\sigma$-algebra on $\mathrm{E}$, and $\mu$ is the probability measure defined by $\mu(\{i,j\})=(1/{\rm deg}_i+1/{\rm deg}_j)/{n}$ for all $\{i,j\}\in \mathrm{E}$. The node selection process can then be seen as a random event in the product probability space $(\Omega,\mathscr{F},\mathbb{P})$, where $\Omega=\mathrm{E}^{\mathbb{N}}=\{\omega=(\omega_0,\omega_1,\dots,): \forall t, \omega_t\in \mathrm{E}\}$, $\mathscr{F}=\mathscr{S}^{\mathbb{N}}$, and $\mathbb{P}$ is the product probability measure (uniquely) defined by: for any finite subset $K\subset \mathbb{N}$, $\mathbb{P}((\omega_t)_{t\in K})=\prod_{t\in K}\mu(\omega_t)$ for any $(\omega_t)_{t\in K}\in \mathrm{E}^{|K|}$. For any $t\in \mathbb{N}$, we define the coordinate mapping $\mathpzc{G}_t:\Omega\to \mathrm{E}$ by $\mathpzc{G}_t(\omega)=\omega_t$, for all $\omega\in \Omega$. Then formally   $\mathpzc{G}_t,\ t=0,1,\ldots$ describe the node pair selection process. We denote $\mathscr{F}_t=\sigma(\mathpzc{G}_0,\ldots,\mathpzc{G}_t)$ as the $\sigma$-algebra capturing the $t+1$ first interactions of the selection process.

After the pair of nodes $\{i,j\}$ have been selected at time $t$, they update their states $x_i(t)$ and $x_j(t)$ according to the sign of the link that they share: if the  link is positive, they update their states by (\ref{positive}); if the link is negative, they update their states by either (\ref{s-f}) or (\ref{r-s-f}). The nodes that are not selected at time $t$ will keep their states unchanged. In this way, $\mathbf{x}(t)$, $t=0,1,\dots$ specifies  a random process over the probability space $(\Omega,\mathscr{F},\mathbb{P})$, and we are interested in the mean, mean-square, and almost sure convergence of  $\mathbf{x}(t)$. { We note that this  signed random gossiping  model has been adopted by \cite{Shi-OR}, and is a special case of the work presented in \cite{Shi-TCNS15,Shi-TCNS17} where switching environments and sign-dependent interaction probabilities were taken into consideration. The current presentation aims for a direct exposure of the same algebra-graphic analysis for random models utilizing the ease from a simplified model. }

\subsection{State Convergence}

For opposing and repelling negative dynamics  models, we present the following results, respectively, for the mean-square and almost sure convergence of $\mathbf{x}(t)$.

\begin{theorem}\label{theorem-sf-random-convergence}
Let $0<\alpha,\beta<1$ and consider  opposing rule (\ref{s-f}) for dynamics over  negative links.
\begin{itemize}
\item[(i)] If $\mathrm{G}$ is  structurally balanced subject to partition $\mathrm{V}=\mathrm{V}_1\mcup \mathrm{V}_2$, then both in the mean-square and almost sure sense there hold
\begin{align}\label{25}
 x_i(t) \rightarrow \big(\sum_{j\in\mathrm{V}_1}x_j(0)-\sum_{j\in\mathrm{V}_2}x_j(0)\big)/n,\ i\in\mathrm{V}_1
 \end{align} and
 \begin{align}\label{26}
 x_i(t) \rightarrow -\big(\sum_{j\in\mathrm{V}_1}x_j(0)-\sum_{j\in\mathrm{V}_2}x_j(0)\big)/n,\ i\in\mathrm{V}_2.
 \end{align}
\item[(ii)] If $\mathrm{G}$ is not  structurally balanced, then $x_i(t)\rightarrow 0$ both in the mean-square and almost sure sense for all $i\in\mathrm{V}$.
\end{itemize}
\end{theorem}

\begin{theorem}\label{theorem-rsf-random-convergence}
  Suppose $\mathrm{G}^+$ is connected and consider repelling rule (\ref{r-s-f}). For any  $0<\alpha<1$, there exists $\beta^\ast(\alpha)>0$  such that  $x_i(t) \rightarrow {\sum_{j=1}^n x_i(0)}/{n}$ both in mean-square and almost surely for all initial value $\mathbf{x}(0)$ if $\beta<\beta^\ast$.
\end{theorem}
{
The almost sure convergence statement of Theorem \ref{theorem-sf-random-convergence} was reported in \cite{Shi-TCNS15}; while the  almost sure convergence statement of Theorem \ref{theorem-rsf-random-convergence} was reported in \cite{Shi-OR}. As the current model gives a stationary graph process, we enjoy the convenience of establishing their proofs using the same mean-square error analysis. }

\subsection{Almost Sure Divergence}

The following results characterize possible   almost sure divergence  of $\mathbf{x}(t)$ caused by large $\beta$ related to the negative links, respectively, for  opposing and repelling  models.

\begin{theorem}\label{theorem-random-asdivergence}
Fix $0<\alpha<1$ with $\alpha\neq 1/2$.
\begin{itemize}
\item[(i)] Suppose both $\mathrm{G}^+$ and $\mathrm{G}^-$ are connected. Then under the opposing  negative dynamics (\ref{s-f}), there exists $\beta_\flat$ such that whenever $\beta>\beta_\flat$, there holds
\begin{align}
\mathbb{P}\Big(\limsup_{t \to \infty} \max_{i\in \mathrm{V}} \big| x_i(t)\big|=\infty\Big)=1
\end{align}
for almost all initial values w.r.t. Lebesgue measure.
\item[(ii)] Suppose  $\mathrm{G}^+$ is connected.  Under the repelling  negative dynamics (\ref{r-s-f}), there exists $\beta_\dag$ such that whenever $\beta>\beta_\dag$, there holds
\begin{align}
\mathbb{P}\Big(\limsup_{t \to \infty} \max_{i,j\in \mathrm{V}} \big| x_i(t)-x_j(t)\big|=\infty\Big)=1
\end{align}
for almost all initial values w.r.t. Lebesgue measure.
\end{itemize}
\end{theorem}

{
The same type of almost sure divergence results can be seen in   \cite{Shi-JSAC13-2,Shi-TCNS15,Shi-OR,Shi-TCNS17} under different random network models. Here  $\alpha\neq 1/2$ is a technical assumption to exclude the case where the positive graph admits finite-time convergence so that the influence of all negative edges is nullified \cite{Shi-OR}. }In fact, for both of the two negative dynamics (\ref{s-f}) and (\ref{r-s-f}), the node states under random node interactions follow a so-called {\em No-Survivor Property} \cite{Shi-OR}, which indicates that every  node states (or relative states) will diverge almost surely if the maximum node states (or relative states) diverges almost surely across the entire network. This property is summarized in the following result.

\begin{theorem}\label{theorem-random-no-survivor}The following statements hold.
\begin{itemize}
\item[(i)] Under the opposing negative dynamics (\ref{s-f}),  there holds for any $k\in\mathrm{V}$ that
\begin{align}
\mathbb{P}\Big(\limsup_{t \to \infty} \big| x_k(t)\big|=\infty \Big| \limsup_{t \to \infty} \max_{i\in \mathrm{V}} \big| x_i(t)\big|=\infty \Big)=1.
\end{align}
\item[(ii)] Suppose   $\mathrm{G}^+$ is connected. Under the repelling  negative dynamics (\ref{r-s-f}),  there holds for any $k\neq m\in\mathrm{V}$ that
\begin{align}
\mathbb{P}\Big(\limsup_{t \to \infty} \big| x_k(t)-x_m(t)\big|=\infty\Big|\limsup_{t \to \infty} \max_{i,j\in \mathrm{V}} \big| x_i(t)-x_j(t)\big|=\infty\Big)=1.
\end{align}
\end{itemize}
\end{theorem}

{
Theorem \ref{theorem-random-no-survivor}.(i) is a special case of Theorem 3 in \cite{Shi-TCNS15}, where general random graph processes are investigated.  Theorem \ref{theorem-random-no-survivor}.(ii) is quoted directly from Theorem 1 in \cite{Shi-OR}. The two statements are established using a sample-path analysis in lights of the Borel-Cantelli lemma (see, e.g., Theorem 2.3.6 in \cite{Durrett-book}). The ``$\limsup$" in the above two theorems can be replaced by ``$\liminf$" and the results continue to hold.}

\subsection{Bounded States for Repelling Dynamics}
Let $A>0$ be a constant and define $\mathscr{P}_A(\cdot)$ by $\mathscr{P}_A(z)=-A, z< -A$, $\mathscr{P}_A(z)=z, z\in [-A,A]$, and $\mathscr{P}_A(z)=A, z>A$.
Define the function $\theta:\mathrm{E}\to \mathbb{R}$ so that $\theta(\{i,j\})=\alpha$ if $\{i,j\}\in\mathrm{E}^+ $ and $\theta(\{i,j\})=-\beta$ if  $\{i,j\}\in\mathrm{E}^-$. Assume that node $i$ interacts with node $j$ at time $t$. We now consider the following node interaction under the repelling  rule:
\begin{align}\label{121}
x_s(t+1)=\mathscr{P}_A\big((1-\theta)x_{s}(t)+ \theta x_{-s}(t)\big), \ s\in\{i,j\}.
\end{align}
Now the node  dynamics in (\ref{121}) become  nonlinear due to the state constraint.   The following result shows that with structural balance of $\mathrm{G}$, state clustering is reached almost surely at the two state boundaries.

\begin{theorem}\label{theorem-strongbalance}
Consider node dynamics (\ref{121}) and let $\alpha\in(0,1/2)$.    Assume that $\mathrm{G}$ is a structurally balanced complete graph under the partition $\mathrm{V}=\mathrm{V}_1\cup  \mathrm{V}_2$.    When  $\beta$ is sufficiently large, for almost all initial values $\mathbf{x}(0)$ w.r.t. Lebesgue measure, there exists  a  binary random variables $l(\mathbf{x}(0))$ taking values in $\{-A,A\}$ such that:
\begin{align}
\mathbb{P}\Big( \lim_{t\rightarrow \infty} x_i(t)=l(\mathbf{x}(0)),  i\in \mathrm{V}_1;\  \lim_{t\rightarrow \infty} x_i(t)=-l(\mathbf{x}(0)),  i\in \mathrm{V}_2\Big)=1.
\end{align}
\end{theorem}

It is interesting to note that the node state clustering results in Theorem \ref{theorem-sf-deterministic} and Theorem \ref{theorem-strongbalance}, for opposing rule and repelling rule, respectively, both rely on structural balance of $\mathrm{G}$. It turns out that when $\mathrm{G}$ is a complete graph, weak structural balance also leads to clustering of  node states.

\begin{theorem}\label{theorem-weakbalance}
Consider node dynamics (\ref{121}) and let $\alpha\in(0,1/2)$.  Assume that $\mathrm{G}$  is a weakly  structurally balanced complete graph under the partition $\mathrm{V}=\mathrm{V}_1\cup  \mathrm{V}_2 \dots \cup\mathrm{V}_m$ with $m\geq 2$.   When  $\beta$ is sufficiently large,  almost sure boundary  clustering is achieved   in the sense that for almost all initial value $\mathbf{x}(0)$ w.r.t. Lebesgue measure,   there are $m$ random variables, $l_1(\mathbf{x}(0)),\dots,l_m(\mathbf{x}(0))$, each of which  taking values in $\{-A,A\}$, such that:
\begin{align}
\mathbb{P}\Big( \lim_{t\rightarrow \infty} x_i(t)=l_j(\mathbf{x}(0)),\   i\in \mathrm{V}_j,\  j=1,\dots,m\Big)=1.\
\end{align}
\end{theorem}

When the positive graph $\mathrm{G}^+$ is connected -- therefore there is no structural balance  --  any node state will touch the two boundaries $-A$ and $A$ an infinite number of times. Recall that the vertex connectivity $\kappa(\mathrm{G})$ of a graph $\mathrm{G}$ is the minimum number of nodes whose removal disconnects $\mathrm{G}$.  The result is summarized below.

\begin{theorem}\label{theorem-ergodic}
Consider node dynamics (\ref{121}) and let $\alpha\in(1/2,1)$.  Assume that $\mathrm{G}$ is a complete graph, and  the positive graph $\mathrm{G}^+$ is connected with $\kappa(\mathrm{G}^+)\geq 2$. When  $\beta$ is sufficiently large, for almost all initial value $\mathbf{x}(0)$ w.r.t. Lebesgue measure,  there holds for all $i\in \mathrm{V}$ that
\begin{align}
\mathbb{P}\Big( \liminf_{t\rightarrow \infty} x_i(t)=-A,\  \limsup_{t\rightarrow \infty} x_i(t)=A\Big)=1.
\end{align}
\end{theorem}

Results of the similar type  as Theorems \ref{theorem-strongbalance}, \ref{theorem-weakbalance} and \ref{theorem-ergodic} were established in \cite{Shi-OR} for a model where asymmetric node updates were also taken into consideration. The current simplified model allows for more direct analysis along the same line of mathematical machinery.    The assumptions on $\mathrm{G}$ being a complete graph and $\alpha$ taking specific range of values are technical assumptions to simplify the analysis, which can be further relaxed. The proofs of  Theorems \ref{theorem-strongbalance}, \ref{theorem-weakbalance} and \ref{theorem-ergodic} are based on stopping time analysis for the process $\mathpzc{G}_t,t=0,1,\dots$ in lights of the Second Borel-Cantelli Lemma, and they have been put in the Appendix.

\section{Conclusions}\label{sec:conclusions}
We have surveyed a few  fundamental results on the convergence properties of  dynamics over signed networks. A unified  approach was provided    in view of generalized  Perron-Frobenius theory, graph theory, and elementary algebraic recursions. The results illustrated that dynamical properties of a network depend crucially on  the  sign structure of the  network links, for both deterministic and random node interactions. Many interesting future research directions emerge naturally  after the connection between such basic  convergence conditions have been clarified. First of all, inverse problems such as estimating characteristics of the annotations of links and nodes   from observations of various network characteristics at a subset of nodes are of primary interest. Typical questions would include re-construction of node initial values,   identification of edge signs,  and test of structural balance  through a perhaps finite sequence of measurements of the node states \cite{Anders-TAC-2013,Materassi-TAC-2013,Mesbahi-TAC-2013,Baras-2014,Anderson-Shi-15}. Another interesting direction would be the investigation of controllability issues related to signed networks along the line of research on network controllability  \cite{Rahmani:2009,Nature,Sundaram_TAC13,Olshevsky2014}. How  sign structure of a network system relates to the network controllability or  structural controllability is still an open problem. Finally,  it is of interest to look into the scenario when the evolving node states  generate feedback to the signs of the network edges. The closed-loop network dynamics  will lead to Krause's type of multi-agent systems where state-dependant interaction structure will inevitably cause high nonlinearity \cite{Hendrickx-TAC-Krause} in the state update at the nodes.

\section*{Appendix}
\subsection*{A. Proof of Theorem \ref{theorem-sf-deterministic-directed}}
The statement (i) again follows directly from  Theorem 2 in \cite{Olfati-Saber-2007} after applying a gauge transformation
$$
z_i(t)=x_i(t), i\in\mathrm{V}_1; \quad z_i(t)=-x_i(t), i\in\mathrm{V}_2.
$$
 We now prove the statement (ii) through a contradiction argument. We proceed in steps.

\medskip

\noindent Step 1. Define $h(t):=\max_{i\in \mathrm{V}} |x_i(t)|$. Observing that (\ref{absolute-stochastic}) continues to hold with a digraph $\mathrm{G}$, we have $h(t+1)\leq h(t)$ for all $t\geq 0$. Consequently, there is a constant $h_\ast(\mathbf{x}(0))>0$ such that $\lim_{t\to \infty}h(t) =h_\ast$ for any initial value $\mathbf{x}(0)$. We only need to consider the case with $h_\ast>0$, and by the definition of $h_\ast$, for any $\epsilon>0$, there exists $T(\epsilon)>0$ such that
\begin{align}\label{11}
|x_i(t)|\leq h_\ast +\epsilon, \ t\geq T.
\end{align}

\medskip

\noindent Step 2. Define $g_i:= \liminf_{t\to \infty}  |x_i(t)|$. In this step, we show $g_i=h_\ast$ for all $i\in\mathrm{V}$. Suppose $g_{i_0}<h_\ast$ for some $i_0\in \mathrm{V}$. By the definition of $g_i$, for any $\epsilon>0$, there always exists $t_1 \geq T$ such that
\begin{align}\label{29}
|x_{i_0}(t_1)|\leq g_{i_0} +\epsilon.
\end{align}

The graph $\mathrm{G}$ is strongly connected. Therefore,  the set $\mathrm{V}^\ast_1:=\big\{j: i_0\in\mathrm{N}_j\big\}$ is nonempty.  Based on  (\ref{11}), (\ref{29}) and the fact that $i_0 \in \mathrm{N}_{i_1}$, we then have
\begin{align}\label{12}
\big|x_{i_1}(t_1+1)\big|&=\bigg|\Big(1-\alpha\big|\mathrm{N}_{i_1}^+\big|-\beta \big|\mathrm{N}_{i_1}^-\big|\Big)x_{i_1}(t)+\alpha \sum_{j\in \mathrm{N}_{i_1}^+}x_j(t)-\beta \sum_{j\in \mathrm{N}_{i_1}^-}x_j(t) \bigg|\nonumber\\
&\leq \Big|1-\alpha\big|\mathrm{N}_{i_1}^+\big|-\beta \big|\mathrm{N}_{i_1}^-\big|\Big|\cdot |x_{i_1}(t)|+\alpha \sum_{j\in \mathrm{N}_{i_1}^+}\big|x_j(t)\big|+\beta \sum_{j\in \mathrm{N}_{i_1}^-}\big|x_j(t)\big|\nonumber\\
&\leq \gamma \big( g_{i_0} +\epsilon\big)+ (1-\gamma)\big(h_\ast +\epsilon\big)\nonumber\\
&= \gamma  g_{i_0} + (1-\gamma)h_\ast +\epsilon
\end{align}
for any $i_1\in \mathrm{V}_1^\ast$, where $\gamma=\min\{\alpha, \beta\}$.

Continuing, we define   $\mathrm{V}_2^\ast:=\big\{j: \exists i_1\in \mathrm{V}_1^\ast,\ i_1\in\mathrm{N}_j\big\}$ as the nodes that have a neighbor in the set $\mathrm{V}_1^\ast$. Again, the set $\mathrm{V}_2^\ast$ is nonempty because the graph $\mathrm{G}$ is strongly connected. Repeating the above analysis we have
\begin{align}\label{13}
\big|x_{i_2}(t_1+2)\big|\leq \gamma^2  g_{i_0} + (1-\gamma^2)h_\ast +\epsilon
\end{align}
for any $i_2\in \mathrm{V}_1^\ast\mcup\mathrm{V}_2^\ast$. This process can be further  recursively carried out, and eventually there must hold
\begin{align}
\big|x_{i}(t_1+n-1)\big|\leq \gamma^{n-1}  g_{i_0} + (1-\gamma^{n-1})h_\ast +\epsilon,\ i\in\mathrm{V}.
\end{align}
Therefore,
\begin{align}
h_\ast\leq \gamma^{n-1}  g_{i_0} + (1-\gamma^{n-1})h_\ast +\epsilon,
\end{align}
or equivalently,
\begin{align}\label{14}
\gamma^{n-1}  \big( h_\ast-g_{i_0}\big)\leq  \epsilon.
\end{align}
This leads to a contradiction if $h_\ast >g_{i_0}$ because $\epsilon$ in (\ref{14}) can be arbitrary.

\medskip

\noindent Step 3. The fact that $g_i=h_\ast$ for all $i\in\mathrm{V}$ immediately leads to $\lim_{t\to \infty} |x_i(t)| =h_\ast$ for all $i\in\mathrm{V}$ since  $\limsup_{t\to \infty}  |x_i(t)|\leq h_\ast$ by the definition of $h_\ast$. It is easy to exclude the case where $\liminf_{t\to \infty} x_i(t) =-h_\ast$ and $\limsup_{t\to \infty} x_i(t) =h_\ast$ for some $i$ directly from the dynamics (\ref{1}). In other words, all node states asymptotically converge. From this point, we can define
 $$
 \mathrm{V}_1:=\big\{i\in\mathrm{V}: \liminf_{t\to \infty} x_i(t) =h_\ast\big\},\  \mathrm{V}_2:=\big\{i\in\mathrm{V}: \liminf_{t\to \infty} x_i(t) =-h_\ast\big\}.
 $$
 It is then clear that the links between $\mathrm{V}_1$ and $\mathrm{V}_2$ can only be negative, and the links inside each subset can only be positive.  This proves that the graph $\mathrm{G}$ is structurally balanced.

We have now concluded the proof.
\hfill$\square$

\subsection*{B. Proof of Theorem \ref{theorem-rsf-deterministic-directed}}
With $\mathrm{G}$ being directed, it still holds that $M_{_\mathrm{G}} \mathbf{1} =\mathbf{1}$ since  $M_{_\mathrm{G}}=I-\alpha{L}_{_{\mathrm{G}^+}}-\beta{L}_{_{\mathrm{G}^-}}^{^{\rm r}}$, where ${L}_{_{\mathrm{G}^+}}\mathbf{1}=0$ and ${L}_{_{\mathrm{G}^-}}^{^{\rm r}}\mathbf{1}=0$ for digraphs $\mathrm{G}^+$ and $\mathrm{G}^-$. Therefore, $1$ is always an eigenvalue of $M_{_\mathrm{G}}$.

Fix $\alpha$ with $0<\alpha<1/\max_{i\in\mathrm{V}} {\rm deg}_i^+$. We can  define the following two functions:
\begin{align}
r(\beta):=\max \Big\{\big|\lambda_i(M_{_\mathrm{G}})\big|:\ \lambda_i(M_{_\mathrm{G}})\in \sigma(M_{_\mathrm{G}})\setminus\{1\}\Big\}
\end{align}
as the largest magnitude of the eigenvalues of $M_{_\mathrm{G}}$ which are not equal to one, and
\begin{align}
\mathbf{q}(\beta):=\big(q_1(\beta) \dots q_n(\beta)\big)
\end{align}
with $\mathbf{q}(\beta)M_{_\mathrm{G}}=\mathbf{q}(\beta)$ and $\sum_{j=1}^n q_j(\beta)=1$.

The following facts stand: (i) $r(0)<1$, and $1$ is a simple eigenvalue of $I-\alpha{L}_{_{\mathrm{G}^+}}$ if $\mathrm{G}^+$ is strongly connected\footnote{In fact, $1$ is a simple eigenvalue of $I-\alpha{L}_{_{\mathrm{G}^+}}$ if $\mathrm{G}^+$ has a directed spanning tree (see, e.g., Proposition 3.8. in \cite{Magnus-book}).}; (ii) $q(0)$ is a positive row vector. Noticing that both $r(\cdot)$ and $q(\cdot)$ are continuous functions,  there exists a sufficiently small $\beta_\ast$ such that both the two facts hold for $\beta<\beta_\ast$, i.e.,  $1$ is a simple eigenvalue of  $M_{_\mathrm{G}}$ with $r(\beta)<1$,  and $\mathbf{q}(\beta)$ is positive. Therefore,  through the Jordan decomposition of $M_{_\mathrm{G}}$, it is easy to see that
$$
\lim_{t\to \infty} M_{_\mathrm{G}}^t=\mathbf{1}\mathbf{q}(\beta),
$$
and this concludes the proof (See the same treatment  applied to continuous-time dynamics in Theorem 3.12, \cite{Magnus-book}).

\subsection*{C. Proof of Theorem \ref{theorem-sf-random-convergence}}
Let $\mathbf{e}_m=(0\dots 1 \dots 0)^\top$ be the $n$-dimensional unit vector whose $m$'th entry is $1$. Under the pair selection process and the opposing rule for negative links, the evolution of the node states can be written as
\begin{align}
\mathbf{x}(t+1)= \mathpzc{W}_t \mathbf{x}(t),
\end{align}
where $ \mathpzc{W}_t$, $t=0,1,\dots$ is an i.i.d. random matrix process. The distribution of $ \mathpzc{W}_t$ is given by
\begin{align}\label{21}
\mathbb{P}\Big( \mathpzc{W}_t=I- \alpha (\mathbf{e}_i-\mathbf{e}_j)(\mathbf{e}_i-\mathbf{e}_j)^\top\Big) =p_{ij},\ \{i,j\}\in\mathrm{E}^+
\end{align}
and
\begin{align}\label{22}
\mathbb{P}\Big( \mathpzc{W}_t=I- \beta (\mathbf{e}_i+\mathbf{e}_j)(\mathbf{e}_i+\mathbf{e}_j)^\top\Big) =p_{ij},\ \{i,j\}\in\mathrm{E}^-.
\end{align}

\medskip

\noindent (i) Let $\mathrm{G}$ be  structurally balanced subject to partition $\mathrm{V}=\mathrm{V}_1\mcup \mathrm{V}_2$. Introduce $J=\mathbf{1}\mathbf{1}^\top/n$,  $K={\rm diag}(k_1,\dots,k_n)$ with $k_i=1$ for $i\in\mathrm{V}_1$ and $k_i=-1$ for $i\in\mathrm{V}_2$, and $\mathbf{k}=(k_1,\dots,k_n)^\top$. Note that, for any realization of $\mathpzc{W}_t$, there holds that
$JK\mathpzc{W}_t=JK$. Thus $K\mathpzc{W}_t K J=JK\mathpzc{W}_t K=J$, which in turn leads to
\begin{align}\label{23}
(I-J) \big(K \mathpzc{W}_tK\big)=\big(K \mathpzc{W}_tK\big)(I-J)
\end{align}

Consider $V(t)=\big\| (I-J)K \mathbf{x}(t)\big\|^2$. Then
\begin{align}\label{30}
\mathbb{E}\Big\{V(t+1)\Big|\mathbf{x}(t) \Big\}&= \mathbb{E}\Big\{\mathbf{x}^\top(t) \mathpzc{W}_tK  (I-J)K \mathpzc{W}_t\mathbf{x}(t)\Big\}\nonumber\\
&\stackrel{a)}{=} \mathbb{E}\Big\{\big(\mathbf{x}^\top(t)K\big) \big(K \mathpzc{W}_tK\big)  (I-J) \big(K \mathpzc{W}_tK\big) \big( K\mathbf{x}(t)\big)\Big\}\nonumber\\
&\stackrel{b)}{=} \mathbb{E}\Big\{\big(\mathbf{x}^\top(t)K (I-J)\big) \big(K \mathpzc{W}_tK\big)  (I-J) \big(K \mathpzc{W}_tK\big) \big(  (I-J)K\mathbf{x}(t)\big)\Big\}\nonumber\\
&\stackrel{c)}{=} \mathbb{E}\Big\{\big(\mathbf{x}^\top(t)K (I-J)\big) \big( K\mathpzc{W}_t^2K- J\big)\big(  (I-J)K\mathbf{x}(t)\big)\Big\}\nonumber\\
&=\big(\mathbf{x}^\top(t)K (I-J)\big) \big( \mathbb{E}\big\{K \mathpzc{W}_t^2K\big\}- J\big)\big(  (I-J)K\mathbf{x}(t)\big)
\end{align}
where $a)$ holds because $K^2=I$, $b)$ is due to the equalities  $(I-J)^2=I-J$ and (\ref{23}), and $c)$ is obtained by applying $JK\mathpzc{W}_t=JK$.

Based on (\ref{21}) and (\ref{22}), we have
\begin{align}\label{24}
{P}_{_{\mathrm{G}}}^\ast:&= \mathbb{E}\big\{ K\mathpzc{W}_t^2K\big\}\nonumber\\
&=\sum_{\{i,j\}\in\mathrm{E}^+} p_{ij}\big(I- 2\alpha(1-\alpha) (\mathbf{e}_i-\mathbf{e}_j)(\mathbf{e}_i-\mathbf{e}_j)^\top\big)+\sum_{\{i,j\}\in\mathrm{E}^-} p_{ij}\big(I- 2\beta(1-\beta) (\mathbf{e}_i-\mathbf{e}_j)(\mathbf{e}_i-\mathbf{e}_j)^\top\big)\nonumber\\
 &=I-2\alpha(1-\alpha){L}_{_{\mathrm{G}^+}}^{^{\rm p}}+2\beta(1-\beta){L}_{_{\mathrm{G}^-}}^{^{\rm pr}},
\end{align}
where ${L}_{_{\mathrm{G}^+}}^{^{\rm p}}$ is the probabilistically weighted Laplacian  of $\mathrm{G}^+$  with $[{L}_{_{\mathrm{G}^+}}^{^{\rm p}}]_{ij}= -p_{ij}$ for $\{i,j\}\in\mathrm{E}^+$, $[{L}_{_{\mathrm{G}^+}}^{^{\rm p}}]_{ij}=0$ for $\{i,j\}\notin\mathrm{E}^+$ with $i\neq j$, and $[{L}_{_{\mathrm{G}^+}}^{^{\rm p}}]_{ii}= \sum_{j\neq i\in\mathrm{N}_i^+}p_{ij}$; ${L}_{_{\mathrm{G}^-}}^{^{\rm pr}}$ is the probabilistically  weighted repelling  Laplacian of $\mathrm{G}^-$ with $[{L}_{_{\mathrm{G}^-}}^{^{\rm pr}}]_{ij}= p_{ij}$ for $\{i,j\}\in\mathrm{E}^-$, $[{L}_{_{\mathrm{G}^-}}^{^{\rm pr}}]_{ij}=0$ for $\{i,j\}\notin\mathrm{E}^-$ with $i\neq j$, and $[{L}_{_{\mathrm{G}^-}}^{^{\rm pr}}]_{ii}= -\sum_{j\neq i\in\mathrm{N}_i^-}p_{ij}$. We note the fact that both  ${L}_{_{\mathrm{G}^+}}^{^{\rm p}}$ and  $-{L}_{_{\mathrm{G}^-}}^{^{\rm pr}}$ are standard weighted Laplacians, and the implication of the properties of their spectrum \cite{Magnus-book} including bounds on their eigenvalues from $\sum_{j}p_{ij}=1$. Also noticing that $\alpha,\beta \in(0,1)$ implies $0<2\alpha(1-\alpha)\leq 1/2$ and  $0<2\beta(1-\beta)\leq 1/2$, the following facts hold.
\begin{itemize}
\item[F1.] $0\leq \lambda_i\big({P}_{_{\mathrm{G}}}^\ast\big)\leq 1$ for all $\lambda_i\big({P}_{_{\mathrm{G}}}^\ast\big)\in\sigma \big({P}_{_{\mathrm{G}}}^\ast\big)$; $1\in \sigma \big({P}_{_{\mathrm{G}}}^\ast\big)$ is a simple eigenvalue  with $\mathbf{1}$ being a corresponding eigenvector.

\item[F2.]   $1$ is an eigenvalue of $\mathbf{1}\mathbf{1}^\top/n$  with multiplicity one and $\mathbf{1}$ is an associated eigenvector; $\mathbf{1}\mathbf{1}^\top/n$ also has zero as its eigenvalue with multiplicity $n-1$.

\item[F3.] ${P}_{_{\mathrm{G}}}^\ast$ and $\mathbf{1}\mathbf{1}^\top$ commute, i.e., ${P}_{_{\mathrm{G}}}^\ast \mathbf{1}\mathbf{1}^\top= \mathbf{1}\mathbf{1}^\top {P}_{_{\mathrm{G}}}^\ast$.
\end{itemize}

Consequently, all eigenvalues of  ${P}_{_{\mathrm{G}}}^\ast-\mathbf{1}\mathbf{1}^\top/n$ is strictly less than one. We can therefore further conclude that
\begin{align}\label{27}
\mathbb{E}\Big\{V(t+1)\big|\mathbf{x}(t) \Big\}\leq \lambda_{\rm max}\big({P}_{_{\mathrm{G}}}^\ast-\mathbf{1}\mathbf{1}^\top/n \big) V(t).
\end{align}
This immediately yields that $\mathbb{E}\big\{V(t)\big\}$ converges to zero, or equivalently, (\ref{25}) and (\ref{26}) hold in the mean-square sense.

Moreover, (\ref{27}) means that $V(t)$ is a supermartingale, which converges to a limit almost surely by the martingale convergence theorem (Theorem 5.2.9, \cite{Durrett-book}). Such a limit must be zero since $0<\lambda_{\rm max}\big({P}_{_{\mathrm{G}}}^\ast-\mathbf{1}\mathbf{1}^\top/n \big)<1$ (which implies, $\mathbb{E}\big\{V(t)\big\}$ converges to zero exponentially), and this concludes that (\ref{25}) and (\ref{26}) hold in the almost sure sense.

\medskip

\noindent(ii) Now we move on to the case where $\mathrm{G}$ is not structurally balanced. Consider instead $V_\ast (t)=\big\|\mathbf{x}(t)\big\|^2$. We have
\begin{align}
\mathbb{E}\Big\{V_\ast(t+1)\big|\mathbf{x}(t) \Big\}
&=\mathbf{x}^\top(t)\mathbb{E}\big\{\mathpzc{W}_t^2\big\}\mathbf{x}(t).
\end{align}
Based on (\ref{21}) and (\ref{22}), we have
\begin{align}\label{24}
{P}_{_{\mathrm{G}}}:&= \mathbb{E}\big\{\mathpzc{W}_t^2\big\}\nonumber\\
&=\sum_{\{i,j\}\in\mathrm{E}^+} p_{ij}\big(I- 2\alpha(1-\alpha) (\mathbf{e}_i-\mathbf{e}_j)(\mathbf{e}_i-\mathbf{e}_j)^\top\big)+\sum_{\{i,j\}\in\mathrm{E}^-} p_{ij}\big(I- 2\beta(1-\beta) (\mathbf{e}_i+\mathbf{e}_j)(\mathbf{e}_i+\mathbf{e}_j)^\top\big)\nonumber\\
 &=I-2\alpha(1-\alpha){L}_{_{\mathrm{G}^+}}^{^{\rm p}}-2\beta(1-\beta){L}_{_{\mathrm{G}^-}}^{^{\rm po}},
\end{align}
where  ${L}_{_{\mathrm{G}^-}}^{^{\rm po}}$ is the probabilistically weighted (opposing) Laplacian of $\mathrm{G}^-$ as a signed graph with $[{L}_{_{\mathrm{G}^-}}^{^{\rm po}}]_{ij}= p_{ij}$ for $\{i,j\}\in\mathrm{E}^-$, $[{L}_{_{\mathrm{G}^-}}^{^{\rm po}}]_{ij}=0$ for $\{i,j\}\notin\mathrm{E}^-$ with $i\neq j$, and $[{L}_{_{\mathrm{G}^-}}^{^{\rm po}}]_{ii}= \sum_{j\neq i\in\mathrm{N}_i^-}p_{ij}$. The main difference between ${W}_{_{\mathrm{G}}}$ and ${P}_{_{\mathrm{G}}}$ lies in the weighted edges in ${P}_{_{\mathrm{G}}}$. Noticing that $\alpha,\beta \in(0,1)$ implies $0<2\alpha(1-\alpha)\leq 1/2$ and  $0<2\beta(1-\beta)\leq 1/2$, there holds
 \begin{align}
 \sum_{j=1}^n \big| [{P}_{_{\mathrm{G}}}]_{ij}\big|=1.
 \end{align}
As discussed previously, the absence of  structural balance of $G$ implies that $$
\lambda_{\rm max} \big( {P}_{_{\mathrm{G}}}\big)<1
$$
as long as $\mathrm{G}$ is a connected graph.
Consequently, we have
\begin{align}
\mathbb{E}\big\{V_\ast(t+1)\big|\mathbf{x}(t) \big\}\leq \lambda_{\rm max} \big( {P}_{_{\mathrm{G}}}\big) V_\ast(t),
\end{align}
which in turn leads to that $\mathbb{E}\{ V_\ast(t)\}$ tends to zero, and that $V_\ast(t)$ goes to zero almost surely from the same analysis applied for $V(t)$. Equivalently, we have proved that $\mathbf{x}(t)$ converges to zero in mean-square and almost surely.

We have now completed the proof of Theorem  \ref{theorem-sf-random-convergence}.

\subsection*{D. Proof of Theorem \ref{theorem-rsf-random-convergence}}
 Let $x_{\rm ave}=\sum_{i\in \mathsf{V}} x_i(0)/n$ be the average of the initial beliefs. We introduce $V_\flat(t)=\sum_{i=1}^n |x_i(t)-x_{\rm ave}|^2= \big\|(I-J)\mathbf{x}(t)\big\|^2$. Similar to (\ref{30}), we have
\begin{align}\label{32}
\mathbb{E} \Big\{ V_\flat(t+1) \Big| \mathbf{x}(t)\Big\}\leq \lambda_{\rm max}\big( \mathbb{E}\{ \mathpzc{W}^2(t)\}-J\big) V_\flat(t).
\end{align}
Under the repelling  rule for negative dynamics, the distribution of $ \mathpzc{W}_t$ is given by
\begin{align}
\mathbb{P}\Big( \mathpzc{W}_t=I- \alpha (\mathbf{e}_i-\mathbf{e}_j)(\mathbf{e}_i-\mathbf{e}_j)^\top\Big) =p_{ij}
\end{align}
if ${\rm Sgn}(\{i,j\})=+$, and
\begin{align}
\mathbb{P}\Big( \mathpzc{W}_t=I+ \beta (\mathbf{e}_i-\mathbf{e}_j)(\mathbf{e}_i-\mathbf{e}_j)^\top\Big) =p_{ij}
\end{align}
if ${\rm Sgn}(\{i,j\})=-$.  As a result, we have
\begin{align}
\mathbb{E} \{\mathpzc{W}^2(t)\}=I- {2\alpha(1-\alpha)} {L}_{_{\mathrm{G}^+}}^{^{\rm p}} - {2\beta(1+\beta )} {L}_{_{\mathrm{G}^-}}^{^{\rm pr}}.
\end{align}
where ${L}_{_{\mathrm{G}^+}}^{^{\rm p}}$ and ${L}_{_{\mathrm{G}^-}}^{^{\rm pr}}$ are defined in (\ref{24}).

Since $\mathrm{G}^+$ is connected, $\lambda_{\rm max}\big (I- {2\alpha(1-\alpha)} {L}_{_{\mathrm{G}^+}}^{^{\rm p}}\big)<1$ noticing $0<\alpha<1$. Consequently, $\lambda_{\rm max}\big(\mathbb{E} \{\mathpzc{W}^2(t)\}-J\big)<1$ for all $\beta$ satisfying
\begin{align}
\beta(1+\beta)< \frac{\lambda_2({L}_{_{\mathrm{G}^+}}^{^{\rm p}})}{\lambda_{\rm max}(-{L}_{_{\mathrm{G}^-}}^{^{\rm pr}})}\alpha(1-\alpha),
\end{align}
where $\lambda_2({L}_{_{\mathrm{G}^+}}^{^{\rm p}})$ is the second smallest eigenvalue of ${L}_{_{\mathrm{G}^+}}^{^{\rm p}}$.
Since $g(\beta)=\beta(1+\beta)$ is nondecreasing, $\lambda_{\rm max}\big(\mathbb{E} \{\mathpzc{W}^2(t)\}-J\big)<1$ for all $0\leq \beta<\beta^\ast$ with
\begin{align}
\beta^\ast:= \sup_{\beta}\Big\{\beta(1+\beta)< \frac{\lambda_2({L}_{_{\mathrm{G}^+}}^{^{\rm p}})}{\lambda_{\rm max}(-{L}_{_{\mathrm{G}^-}}^{^{\rm pr}})}\alpha(1-\alpha)\Big\}.
\end{align}
Applying the same analysis that is used for $V(t)$ and $V_\ast(t)$, for any $0\leq \beta<\beta^\ast$ and from (\ref{32}), there hold that $\mathbb{E} \big\{V_\flat(t)\big\}$ converges to zero, and that $V_\flat(t)$ tends to zero almost surely. This completes the proof of Theorem \ref{theorem-rsf-random-convergence}.

\subsection*{E. Proof of Theorem \ref{theorem-random-asdivergence}}

\noindent (i) Define $ h (t):=\max_{i\in\mathrm{V}}\big| x_i(t)\big|$. The proof is based on the following lemma.

\begin{lemma}\label{lemma-lowerbound}
Let $\alpha\neq 1/2\in(0,1)$ and $\beta\geq 3$. Then $\big\{  h (t+1)\geq \min\{\big|2\alpha-1\big|, 1/2\} h (t)\big\}$ is a sure event.
\end{lemma}
{\it Proof.} We discuss two cases, respectively.
\begin{itemize}
\item[C1.] Suppose a pair of nodes $\{i,j\}$ sharing a positive link is selected at time $t$. If both $\big| x_i(t)\big|< h (t)$ and $\big| x_j(t)\big|< h (t)$ hold, then $ h (t+1)\geq h (t)$. Therefore, we assume without loss of generality that $\big| x_i(t)\big|= h (t)$. This leads to two scenarios.
    \begin{itemize}
    \item[(a)] Let $0<\alpha<1/2$. Then
    \begin{align}\label{41}
    \big|x_i(t+1) \big|=\big|(1-\alpha)x_i(t)+\alpha x_j(t) \big|\geq (1-\alpha)\big|x_i(t)\big|-\alpha \big|x_j(t)\big|\geq (1-2\alpha) h (t).
    \end{align}
      \item[(b)] Let $1/2<\alpha<1$. Then
    \begin{align}\label{42}
    \big|x_j(t+1) \big|=\big|(1-\alpha)x_j(t)+\alpha x_i(t) \big|\geq \alpha\big|x_i(t)\big|-(1-\alpha) \big|x_j(t)\big|\geq (2\alpha-1) h (t).
    \end{align}
    \end{itemize}
    We see (\ref{41}) and (\ref{42}) lead to  $ h (t+1)\geq \big|2\alpha-1\big | h (t)$.

   \item[C2.]  Suppose a pair of nodes $\{i,j\}$ sharing a negative link is selected at time $t$.  Again we assume without loss of generality that $\big| x_i(t)\big|= h (t)$. We define $y_i(t)=x_i(t)$ and $y_j(t)=-x_j(t)$. Then the update of $y_i(t)$ and $y_j(t)$ is described by
       \begin{align}\label{45}
       \begin{aligned}
       & y_i(t+1)= y_i(t) +\beta \big(y_j(t)-y_i(t)\big) \\
       & y_j(t+1)= y_j(t) +\beta \big(y_i(t)-y_j(t)\big)
       \end{aligned}
       \end{align}
       \begin{itemize}
     \item[(a)]  If $\big| y_j(t)\big|\geq   h (t)/2$, we see obviously from (\ref{45}) that
       \begin{align}\label{70}
        h (t+1)\geq\big| y_j(t+1)\big|\geq   h (t)/2
       \end{align}
        if $y_i(t)$ and $y_j(t)$ have the same sign. Otherwise without loss of generality let $y_i(t)>0$ and $y_j(t)<0$. Then from (\ref{45})
       \begin{align}\label{71}
      \big| y_i(t+1)\big|&= \big|y_i(t) +\beta \big(y_j(t)-y_i(t)\big)\big| \nonumber\\
      &\geq \beta \big|y_j(t)-y_i(t)\big| -\big|y_i(t) \big|\nonumber\\
      &\geq \frac{3}{2}\beta  h (t)- h (t)\nonumber\\
      &\geq  h (t)/2
       \end{align}
       for $\beta\geq 1$.
       \item[(b)] If $\big| y_j(t)\big|<   h (t)/2$, then there holds for $\beta\geq 3$ that
       \begin{align}\label{72}
          \big| y_i(t+1)\big|&= \big|(1-\beta)y_i(t) +\beta y_j(t)\big| \nonumber\\
      &\geq (\beta-1) \big|y_i(t)\big| -\beta\big|y_j(t) \big|\nonumber\\
      &\geq \big(\frac{1}{2}\beta-1\big)   h (t)\nonumber\\
      &\geq   h (t)/2
       \end{align}
       \end{itemize}
       We see (\ref{70}), (\ref{71}), and (\ref{72}) lead to $ h (t+1)\geq  h (t)/2$ if $\beta\geq 3$.
\end{itemize}
We have now proved the desired lemma. \hfill$\square$

With Lemma \ref{lemma-lowerbound} serves as the same role as the Lemma 5 in \cite{Shi-TCNS15}, the desired conclusion follows from the same argument in view of the strong law of large numbers as the proof of Proposition 1 of \cite{Shi-TCNS15}. We therefore omit the remaining  details.

\noindent (ii) The result comes from Theorem 3 in \cite{Shi-OR}. We therefore refer to the proof therein, which is also based on the strong law of large numbers.

\subsection*{F. Proof of Theorem \ref{theorem-strongbalance}}
We quote the following lemma, Lemma 7 in \cite{Shi-OR}. Note that the proof of Lemma 7 in \cite{Shi-OR} does not rely on the asymmetric node updates, and therefore the lemma continues to hold for  (\ref{121}).
\begin{lemma}\label{lembounddivergence}
Fix $\alpha\in (0, 1) $ with $\alpha\neq 1/2$. For the dynamics (\ref{121}) with the random pair selection process, there exists $\beta^\diamond(\alpha) >0$ such that  $$
\mathbb{P}\Big(\limsup_{t\rightarrow \infty} \max_{i,j\in\mathrm{V}}\big|x_i(t)-x_j(t) \big|=2A\Big)=1
$$ for almost all initial beliefs  if  $ \beta>\beta^\diamond$.
\end{lemma}

We  establish another technical lemma.
\begin{lemma}\label{lem-clustering}
Fix $\alpha\in (0, 1/2)$ and $\beta\geq 1/\alpha$. Consider  the dynamics (\ref{121}) with the random pair selection process. Assume that $\mathrm{G}$ is a structurally balanced complete graph under the partition $\mathrm{V}=\mathrm{V}_1\cup  \mathrm{V}_2$. If there are $i_1\in \mathrm{V}_1$, $j_1\in \mathrm{V}_2$ and $t\geq 0$ with  $x_{i_1}(t)=-A$ and $x_{j_1}(t)=A$, then for $Z=3(n-2)$, there exists and a sequence of node pair realizations, $\mathpzc{G}_{t+s}(\omega)$ for $s=0,1,\dots,Z-1$ under which there holds
\begin{align}
x_i({t+Z})(\omega)=-A, i\in \mathrm{V}_1;\ x_i(t+Z)(\omega)=A, i\in\mathrm{V}_2.
\end{align}
\end{lemma}
{\it Proof.} We recursively construct such sequence of node pair realizations  $\mathpzc{G}_{t+s}(\omega)$ for $s=0,1,\dots,Z-1$. Without loss of generality we let $\mathrm{V}_1$ contain at least two nodes.

Take $i_2 \neq i_1 \in \mathrm{V}_1$ and let
\begin{align}
\mathpzc{G}_{t}(\omega)=\{i_1,i_2\},\mathpzc{G}_{t+1}(\omega)=\{j_1,i_1\}, \mathpzc{G}_{t+2}(\omega)=\{j_1,i_2\}.
\end{align}
Now we investigate the outcome of the above pair selection process. Since $i_1,i_2\in\mathrm{V}_1$, they share a positive link whose interaction is defined by (\ref{positive}). Consequently, we conclude from $x_{i_1}(t)=-A$ and $\alpha\in(0,1/2)$ that
\begin{align}
x_{i_1}(t+1)(\omega)\leq 0, \ x_{i_2}(t+1)(\omega)\leq (1-2\alpha)A.
\end{align}
Further, since $\beta\geq 1/\alpha \geq  2$ and $x_{j_1}(t)=A$,  with the chosen $\mathpzc{G}_{t+1}(\omega)$ we have 
\begin{align}
x_{i_2}(t+2)(\omega)\leq (1-2\alpha)A, \ x_{i_1}(t+2)(\omega)=-A, \ x_{j_1}(t+2)(\omega)= A.
\end{align}
Finally, noticing the fact that $\beta\geq 1/\alpha$ there holds
\begin{align}
x_{i_1}(t+3)(\omega)=-A,\ x_{i_2}(t+3)(\omega)=-A, \ x_{j_1}(t+3)(\omega)= A.
\end{align}

Next, we recursively apply the above pair selections for other nodes in $\mathrm{V}_1$ and then we get $x_{j_1}(t+3n_1)(\omega)= A$ and
\begin{align}
x_{i}(t+3n_1)(\omega)=-A, \ i\in\mathrm{V}_1
\end{align}
with $n_1= |\mathrm{V}_1|-1$.

Finally, we repeat the same pair selection process for nodes in $\mathrm{V}_2$. This will yield
\begin{align}
x_{i}(t+3(n-2))(\omega)=-A, \ i\in\mathrm{V}_1; \quad x_{i}(t+3(n-2))(\omega)=A, \ i\in\mathrm{V}_2.
\end{align}
This proved the desired lemma. \hfill$\square$

We now have the necessary tools in hands for the proof of Theorem \ref{theorem-strongbalance}. By Lemma \ref{lembounddivergence}, there are two nodes $i_\ast$ and $j_\ast$ such that  with probability one,
\begin{align}\label{60}
\limsup_{t\rightarrow \infty} \big|x_{i_\ast}(t)-x_{j_\ast}(t) \big|=2A.
\end{align}
We  define
$$
\mathpzc{T}_1:\inf_{t\geq 0} \big|x_{i_\ast}(t) -x_{j_\ast}(t)\big|\geq A
$$
and then recursively define
$$
\mathpzc{T}_{m+1}:\inf_{t\geq \mathpzc{T}_m+1} \big|x_{i_\ast}(t) -x_{j_\ast}(t)\big|\geq A
$$
for $m=2,3,\dots$. Evidently they form a sequence of stopping times \cite{Durrett-book} in the random node pair process $\mathpzc{G}_t, t=0,1,\dots$. From the fact that (\ref{60}) holds with probability one, $\mathpzc{T}_{m}$ is almost surely finite for any $m=1,2,\dots$.

There will be two cases.

\begin{itemize}
\item[C1.]  Let $i_\ast$ and $j_\ast$ belong to different subgroups, say, $i_\ast\in \mathrm{V}_1$ and  $j_\ast\in \mathrm{V}_2$. Then by selecting $\{i_\ast,j_\ast\}$ at time $\mathpzc{T}_m$, we have
\begin{align}\label{61}
x_{i_1}(\mathpzc{T}_m+1)=-A,\ x_{j_1}(\mathpzc{T}_m+1)=A,
\end{align}
where $i_1$ and $j_1$ are from the set $\{i_\ast,j_\ast\}$ sharing a negative link. Let $i_1\in \mathrm{V}^{i_1}$  and $i_2\in \mathrm{V}^{i_2}$, where each $\mathrm{V}^{i_1}$ and $\mathrm{V}^{i_2}$ is either $\mathrm{V}_1$ or $\mathrm{V}_2$. Then Lemma \ref{lem-clustering} suggests from (\ref{61}) that
\begin{align}\label{64}
\mathbb{P} \Big(x_i\big(\mathpzc{T}_m+Z+1\big)=-A, i\in \mathrm{V}^{i_1};\ x_i\big(\mathpzc{T}_m+Z+1\big)=A, i\in\mathrm{V}^{i_2}\Big) \geq \big(\min_{\{i,j\}\in \mathrm{E}}p_{ij}\big)^{Z+1}.
\end{align}
Note that, since the $\mathpzc{T}_{m}$ are stopping times of $\mathpzc{G}_t,t=0,1,\dots$, by strong Markov property we can invoke the  second Borel-Cantelli Lemma (e.g., Theorem 2.3.6 in \cite{Durrett-book}) to conclude from (\ref{64}) that almost surely, there is $m_0\in \mathbb{Z}^+$ such that
$$
x_i\big(\mathpzc{T}_{m_0}+Z+1\big)=-A, i\in \mathrm{V}^{i_1};\ x_i\big(\mathpzc{T}_{m_0}+Z+1\big)=A, i\in\mathrm{V}^{i_2},
$$
and therefore
$$
x_i(t)=-A, i\in \mathrm{V}^{i_1};\ x_i(t)=A, i\in\mathrm{V}^{i_2}
$$
for all $t\geq\mathpzc{T}_{m_0}+Z+1$ from the structure of the dynamics.

\item[C2.]  Let $i_\ast$ and $j_\ast$ belong to the same subgroup, say, $\mathrm{V}_1$. There must be another node $k_\ast\in \mathrm{V}_2$ such that either $\big|x_{i_\ast}(\mathpzc{T}_m) -x_{k_\ast}(\mathpzc{T}_m)\big|\geq A/2$ or $\big|x_{j_\ast}(\mathpzc{T}_m) -x_{k_\ast}(\mathpzc{T}_m)\big|\geq A/2$. No matter which case it is by selecting the corresponding pair $\{i_\ast,k_\ast\}$ or $\{j_\ast,k_\ast\}$ for time $\mathpzc{T}_m$, we obtain two nodes $i_1$ ($=i_\ast\ \mbox{or}\ j_\ast$) and $j_1$ ($=k_\ast$) so that
\begin{align}
x_{i_1}(\mathpzc{T}_m+1)=-A,\ x_{j_1}(\mathpzc{T}_m+1)=A.
\end{align}
Consequently, this case also ends up with condition (\ref{61}) and therefore rest of treatment remains the same.
\end{itemize}

We have now completed the proof of Theorem \ref{theorem-strongbalance}.

\subsection*{G. Proof of Theorem \ref{theorem-weakbalance}}

Following  Lemma \ref{lem-clustering}, another lemma can be established.
\begin{lemma}\label{lem-clustering-weak}
Fix $\alpha\in (0, 1/2)$ and $\beta\geq 1/\alpha$. Consider  the dynamics (\ref{121}) with the random pair selection process. Let $\mathrm{G}$  be  a weakly  structurally balanced complete graph under the partition $\mathrm{V}=\mathrm{V}_1\cup  \mathrm{V}_2 \dots \cup\mathrm{V}_m$ for $m\geq 2$. If there are $i_1\in \mathrm{V}_1$, $j_1\in \mathrm{V}_2$ and $t\geq 0$ with  $x_{i_1}(t)=-A$ and $x_{j_1}(t)=A$, then for $Z=3n-2m-2$, there exists and a sequence of node pair realizations, $\mathpzc{G}_{t+s}(\omega)$ for $s=0,1,\dots,Z-1$ under which there holds
\begin{align}
x_i({t+Z})(\omega)=-A, i\in \mathrm{V}_1;\ x_i(t+Z)(\omega)=A, i\in\mathrm{V}_2;\ x_i({t+Z})(\omega)=\mathcal{I}_0A, i\in \mathrm{V}_m, m\geq3,
\end{align}
where $\mathcal{I}_0$ takes value from $\{-1,1\}$ relying on $\mathbf{x}(t)$.
\end{lemma}
{\it Proof.} First of all we apply the node pair selection process in the proof of Lemma \ref{lem-clustering} and get with $Z_1=3(|\mathrm{V}_1|+|\mathrm{V}_2|-2)$ that
\begin{align}
x_i({t+Z_1})(\omega)=-A, i\in \mathrm{V}_1;\ x_i(t+Z_1)(\omega)=A, i\in\mathrm{V}_2.
\end{align}

Now take $k_1\in \mathrm{V}_3$. Either $x_{k_1}(t)=x_{k_1}(t+Z_1)<0$ or $x_{k_1}(t)=x_{k_1}(t+Z_1)\geq 0$ must hold. If $x_{k_1}(t+Z_1)<0$ then letting  $\mathpzc{G}_{t+Z_1}=\{k_1,j_1\}$ we have
$
x_{k_1}(t+Z_1+1)=-A,\ x_{j_1}(t+Z_1+1)=A
$. Applying the proof of Lemma \ref{lem-clustering} to $\mathrm{V}_3$, there is a sequence of node pairs leading to
$$
x_{i}(t+Z_1+3|\mathrm{V}_3|-2)=-A,\ i\in\mathrm{V}_3.
$$ Similarly, the other case with $x_{k_1}(t)=x_{k_1}(t+Z_1)\geq 0$ leads to
$$
x_{i}(t+Z_1+3|\mathrm{V}_3|-2)=A,\ i\in\mathrm{V}_3.
$$

The process can be recursively carried out to the rest of the nodes. The whole process counts $3(n-m)+m-2 =3n-2m-2$ node pairs. The desired conclusion holds. \hfill$\square$

The same argument based on stopping times of $\mathpzc{G}_t$ and the second Borel-Cantelli Lemma in the proof of Theorem \ref{theorem-strongbalance} can now be applied to the weakly structural balance case with the help of Lemma \ref{lem-clustering-weak}, and then Theorem \ref{theorem-weakbalance} holds.

\subsection*{H. Proof of Theorem \ref{theorem-ergodic}}
The proof is based on the following lemma.
\begin{lemma}\label{lem-ergodic}
Fix $\alpha\in (1/2,1)$ and $\beta\geq  {2}/({2\alpha -1})$. Consider  the dynamics (\ref{121}) with the random pair selection process. Let $\mathrm{G}$  be  the complete graph   with  $\kappa(\mathrm{G}^+)\geq 2$. Suppose for time $t$ there are $i_1, j_1\in \mathrm{V}$ with  $x_{i_1}(t)=-A$ and $x_{j_1}(t)=A$. Then for any $\epsilon\in[0,(2\alpha-1)A/2\alpha]$ and any $i_\star\in\mathrm{V}$, the following statements hold. 

(i) There exist an integer  $Z(\epsilon)$ and a sequence of node pair realizations, $\mathpzc{G}_{t+s}(\omega)$ for $s=0,1,\dots,Z-1$ under which $x_{i_\star}({t+Z})(\omega)\leq -A+\epsilon$.

(ii) There exist an integer  $Z(\epsilon)$ and a sequence of node pair realizations,  $\mathpzc{G}_{t+s}(\omega)$ for $s=0,1,\dots,Z-1$ under which $x_{i_\star}({t+Z})(\omega)\geq A-\epsilon$. 
\end{lemma}
{\it Proof.} From our standing assumption the negative graph $\mathrm{G}^-$ contains at least one edge. Let $k_\ast, m_\ast \in\mathrm{V}$ share a negative link. We assume the two nodes $i_1, j_1\in \mathrm{V}$ defined in the lemma are different from $k_\ast, m_\ast$, for the ease of the presentation. We can then analyze  all possible sign patterns among the four nodes $i_1,j_1,k_\ast,m_\ast$. We just present the analysis for the  case with
 $$
 \{i_1,k_\ast\}\in\mathrm{E}^+, \{i_1,m_\ast\}\in\mathrm{E}^+, \{j_1,k_\ast\}\in\mathrm{E}^+, \{j_1,m_\ast\}\in\mathrm{E}^+.
 $$
 The other cases are indeed simpler and can be studied via similar techniques.
 
Without loss of generality we let $x_{m_\ast}(t)\geq x_{k_\ast}(t)$. First of all we select $\mathpzc{G}_{t}=\{i_1,k_\ast\}$ and $\mathpzc{G}_{t+1}=\{j_1,m_\ast\}$. It is then straightforward to verify that
$$
x_{m_\ast}(t+2)\geq x_{k_\ast}(t+2)+2\alpha A.
$$
By selecting $\mathpzc{G}_{t+2}=\{m_\ast, k_\ast\}$ we know from $\beta\geq  {2}/({2\alpha-1}) \geq 1/\alpha$ that
$$
x_{k_\ast}(t+3)=-A, \ x_{m_\ast}(t+3)=A.
$$
There will be two cases. 

\begin{itemize}
\item [(a)] Let $i_\star\notin \{m_\ast,k_\ast\}$. Noting that $\kappa(\mathrm{G}^+)\geq 2$, there will be a path connecting to $k_\ast$ from $i_\star$ without passing through $m_\ast$ in $\mathrm{G}^+$. It is then obvious that we can select a  finite number $Z_1$ of links which alternate between $\{m_\ast,k_\ast\}$ and the edges over that path so that 
$
x_{i_\star}(t+3+Z_1)\geq A-\epsilon.
$ Here $Z_1$ depends only on $\alpha$ and $n$. Similarly, there is also a path connecting to $m_\ast$ from $i_\star$ without passing through $k_\ast$ in $\mathrm{G}^+$, based on which we can select realizations of node pairs guaranteeing $
x_{i_\star}(t+3+Z_1)\leq -A+\epsilon.
$

\item[(b)] Let $i_\star\in\{m_\ast,k_\ast\}$. We only need to show that we can select pair realizations so that $x_{m_\ast}$ can get close to $-A$, and $x_{k_\ast}$ gets close to $A$ after $t+3$. Since $\mathrm{G}^+$ is connected, either $m_\ast$ or $k_\ast$ has at least one positive neighbors. We for the moment assume $m'$ is a positive neighbor of $m_\ast$ and $k'$ is a positive neighbor of $k_\ast$ with $m'\neq k'$.  Then from Part (a) we can select $Z_2$ node pairs so that 
$$
x_{m'}(t+3+Z_2)\leq -A+\epsilon, \ x_{k'}(t+3+Z_2)\geq A-\epsilon.
$$
Thus, selecting $\{m',m_\ast\}$ and $\{k',k_\ast\}$ for the next two time instances leads to 
$$
x_{m_\ast}(t+5+Z_2)\leq (1-2\alpha)A+ \alpha \epsilon \leq (1-2\alpha)A/2, \ x_{k_\ast}(t+5+Z_2)\geq (2\alpha-1)A- \alpha \epsilon \geq (2\alpha-1)A/2.
$$
Selecting the negative edge $\{m_\ast,k_\ast\}$ for $t+5+Z_2$ implies $
x_{m_\ast}(t+6+Z_2)=-A, \ x_{k_\ast}(t+6+Z_2)=A
$
for $\beta\geq {2}/({2\alpha-1})$. The case with $m'=k'$ can be dealt with by a similar treatment leading to the same conclusion. 
\end{itemize}
  This concludes the proof of the lemma. \hfill$\square$

In view of Lemma \ref{lembounddivergence} and Lemma \ref{lem-ergodic},  the desired theorem is,  again,  a consequence of the second Borel-Cantelli Lemma.

\end{document}